\documentclass{elsarticle}

\usepackage{url}
\usepackage{hyperref}
\usepackage{romannum}
\usepackage[version-1-compatibility,separate-uncertainty=true]{siunitx}
\usepackage{graphicx}
\usepackage{subfig}
\usepackage{lineno}
\graphicspath{{Figs/}}

\begin{document}

\begin{frontmatter}

  %\title{Production and validation results of Korean large-sized GEM foils for the CMS phase-2 upgrade of muon spectrometer}
  \title{Production and validation of industrially produced large-sized GEM foils for the Phase-2 upgrade of the CMS muon spectrometer} 
  \author[n]{M.~Abbas} %%done
\author[s]{M.~Abbrescia} %%done
\author[i,k]{H.~Abdalla} %%done
\author[i,l]{A.~Abdelalim} %%done
\author[j,w]{S.~AbuZeid}  %%done
\author[ag]{D.~Aebi}  %%done
\author[e]{A.~Agapitos}  %%done
\author[ac]{A.~Ahmad} %%done
\author[q]{A.~Ahmed}  %%done
\author[ac]{W.~Ahmed}  %%done
% ...
%%done

\fntext[note1]{Present address: University of Florida, Gainesville, USA}
\fntext[note2]{Present address: University of Notre Dame, Notre Dame, USA}
\fntext[note3]{Present address: Max Planck Institut für Physik, München, Germany}
\fntext[note4]{Present address: Korea Astronomy and Space Science Institute, Daejeon, Republic of Korea}
\fntext[note5]{Present address: Argonne National Laboratory, Lemont, USA}
\fntext[note6]{Present address: Boston University, Boston, USA}
\fntext[note7]{Present address: Karlsruhe Institute of Technology, Karlsruhe, Germany}

\cortext[cor1]{Corresponding author.}
%\author{\\on behalf of the CMS Muon Group}

% ...
\address[e]{Peking University, Beijing, China} %
% ...
\address[i]{Academy of Scientific Research and Technology - ENHEP, Cairo, Egypt} %
\address[j]{Ain Shams University, Cairo, Egypt}
\address[k]{Cairo University, Cairo, Egypt}
\address[l]{Helwan University, also at Zewail City of Science and Technology, Cairo, Egypt}
% ...
\address[n]{Karlsruhe Institute of Technology, Karlsruhe, Germany}
% ...
\address[q]{Delhi University, Delhi, India}
% ...
\address[s]{Politecnico di Bari, Universit\`{a} di Bari and INFN Sezione di Bari, Bari, Italy}%
% ...
\address[w]{Universit\`{a} di Pavia and INFN Sezione di Pavia, Pavia, Italy} %
% ...
\address[ac]{National Center for Physics, Islamabad, Pakistan}
% ...
\address[ag]{Texas A$\&$M University, College Station, USA}
% ...

  \begin{abstract}
  The upgrade of the CMS detector for the high luminosity LHC (HL-LHC) will include gas electron multiplier (GEM) detectors in the end-cap muon spectrometer.
  %Due to the limited supply of large area GEM detectors, the  Korean CMS (KCMS) collaboration has formed a consortium with Mecaro Co., Ltd. to serve as a supplier of GEM foils with area of approximately \SI{0.6}{\meter\squared}.
  Due to the limited supply of large area GEM detectors, the  Korean CMS (KCMS) collaboration had formed a consortium with Mecaro Co., Ltd. to serve as a supplier of GEM foils with area of approximately \SI{0.6}{\meter\squared}.
  %\GEoneoneArea.
  The consortium has developed a double-mask etching technique for production of these large-sized GEM foils.
    This article describes the production, quality control, and quality assessment (QA/QC) procedures and the mass production status for the GEM foils. 
  Validation procedures indicate that the structure of the Korean foils are in the designed range.
  %The detectors were optimized to satisfy the %requirements of the HL-LHC in terms of the %effective gain, response uniformity, rate %capability, discharge probability, and hardness %against discharges.
  Detectors employing the Korean foils satisfy the requirements of the HL-LHC in terms of the effective gain, response uniformity, rate capability, discharge probability, and hardness against discharges.
  No aging phenomena were observed with a charge collection of \SI{82}{\milli\coulomb\per\centi\meter\squared}.
  %After these performance validation, GEM foil %mass production for the CMS upgrades is being %carried out.
  %
Mass production of KCMS GEM foils is currently in progress. 
%
  %This article describes the production, quality control, and quality assessment (QA/QC) procedures for the large-sized GEM foils for the CMS end-cap muon chambers.
\end{abstract}


\end{frontmatter}

\section{Introduction}
Since the concept of a gas electron multiplier (GEM) %technology 
was introduced \cite{SAULI1997531}, the technology has garnered considerable attention from the experimental particle and nuclear physics communities.
Owing to its substantially high flux capability, significant hardness to radiation, and high position resolution, GEM technology has been widely applied and adopted for next generation tracking devices in several experiments, including the CMS experiment.

To take advantage of the increased luminosity of the HL-LHC compared to that of the LHC \cite{Apollinari:2284929}, upgrades of the CMS experiment are planned to enhance performance under HL-LHC operating conditions.
%
%Because luminosity upgrades of the LHC (HL-LHC) are ongoing %\cite{Apollinari:2284929}, upgrades for CMS detectors are %also planned to maintain or even improve CMS performances %with HL-LHC operating conditions.
%
The upgrades include three stations of GEM detectors for the endcap muon system \cite{Colaleo:2021453, Collaboration:2283189}.
These stations, in order of their distance from the interaction point, are called ME0, GE1/1, and GE2/1.
The installation of the GE1/1 stations is complete and commissioning is ongoing. Production and assembly of GE2/1 detectors has just begun and will be followed by production for ME0. The GE2/1 (ME0) stations are scheduled to be installed during the LHC year end technical stops 2023-2024, and 2024-2025 (LHC long shutdown 3, 2026-2028) \cite{LHC_Operation_Schedule, Lamont_2022}.
%\cite{Schmidt_2016}. 

%Should the information in this paragraph be moved later or eliminated?
% PK: I added sentences motivating the mention of long/short and M1-M8.
The GE1/1 station is comprised of two types of detector modules, long and short, to accommodate existing support structures while maximizing acceptance. A short-type module was used for the study presented here. To cover a large area with uniform detector properties, the GE2/1 detectors are comprised of eight modules, labeled M1 to M8. An M2-type module is being used for the aging study described at the end of Subsection \ref{subsec::detector_validation}.
% use correct description of M1-M8 sectors

%The Korean CMS group (KCMS) and \MECARO, a Korean company producing components and materials for semiconductor production \cite{Mecaro}, have formed a consortium that designates the KCMS group as the second supplier of large-sized GEM foils.
%The Korean CMS group (KCMS) and \MECARO, a Korean company producing components and materials for semiconductor production \cite{Mecaro}, had formed a consortium that designates the KCMS group as the second supplier of large-sized GEM foils.
The Korean CMS group (KCMS) is designated as a second supplier of large-sized GEM foils for the CMS upgrades. 
%The production of large area GEM detectors with stringent and highly application specific properties has posed a problem for industrial production.
For a long time, the CERN micro pattern technologies (MPT) workshop was the sole supplier of large-sized GEM foils.
%However, CERN is a research organization, and not a facility for mass production.
%CERN MPT could not solely satisfy the increasing demand for GEM foils for the entire experimental physics community.
This motivated the CMS experiment to find a second supplier of large-sized GEM foils in the industrial sector.
For this purpose, the KCMS had formed a consortium with MECARO Co., Ltd., a Korean company producing components and materials for semiconductor production \cite{Mecaro}.
The objective of the consortium was to contribute to the CMS upgrade, as well as other experiments, by supplying large-sized GEM foils.
Korean GEM foils will be produced for and deployed in the GE2/1 and ME0 detectors.
\section{GEM foil production}

%GEM foils are flexible Cu clad laminates (FCCLs) typically with %micro holes of \SI{50}{\micro\meter} thick polyimide (PI) coated %on both sides with \numrange[range-phrase = %--]{3}{5}\,\si{\micro\meter} thick Cu.
%The conventional diameter of the Cu (PI) holes is \num{70 \pm 5} %(\num{55 \pm 5})\,\si{\micro\meter} with a pitch of %\SI{140}{\micro\meter}.

GEM foils are made from flexible copper-clad laminate (FCCL) consisting of \SI{50}{\micro\meter} thick polyimide (PI) coated on both sides with \numrange[range-phrase = --]{3}{5}\,\si{\micro\meter} thick Cu.
The foils are perforated by symmetric bi-conical holes with diameter \num{70 \pm 5} \si{\micro\meter} (Cu) and \num{55 \pm 5}\,\si{\micro\meter} (PI). The holes are separated with a pitch of \SI{140}{\micro\meter}.

%The consortium produces large-sized GEM foils using a double-mask %technique.
%It is the world's first and the sole large-sized GEM foil supplier %that adopts the double-mask technique.
%The technique is selected for faster production because it %facilitates a production process that is significantly simpler %than the single-mask technique \cite{Pinto_2009}.
%With the double-mask technique and semi-automated machinery, a %faster production rate is achievable.
%The current production rate is approximately 10 foils per week and %is expected to increase as production experience is accumulated.
%The shape of the developed hole is symmetrically bi-conical.

The consortium is the world's first and sole producer of large-sized GEM foils using a double-mask technique. This technique, employing semi-automated machinery, is significantly simpler and faster than the previously used single-mask technique \cite{Pinto_2009}. 
%The current production rate is approximately 10 foils per week and is expected to increase as production experience is accumulated.

%The alignment of the top and bottom masks is extremely crucial in developing GEM foils with proper geometry as the double-mask technique is %chosen.
%The permissible misalignment is less than \SI{7}{\micro\meter}.
%Fig.~\ref{fig:exposure} presents the large-sized bipolar ultraviolet (UV) exposure machine.
%Before irradiating with UV rays to implant a pattern through the masks onto dry film photoresists (DFRs) that are laminated on both sides of %the FCCL, this machine aligns the masks.
%The misalignment was verified to approximately \SI{5}{\micro\meter}.

\begin{figure}[hbt]
  \centering
  \includegraphics[width=0.6\columnwidth]{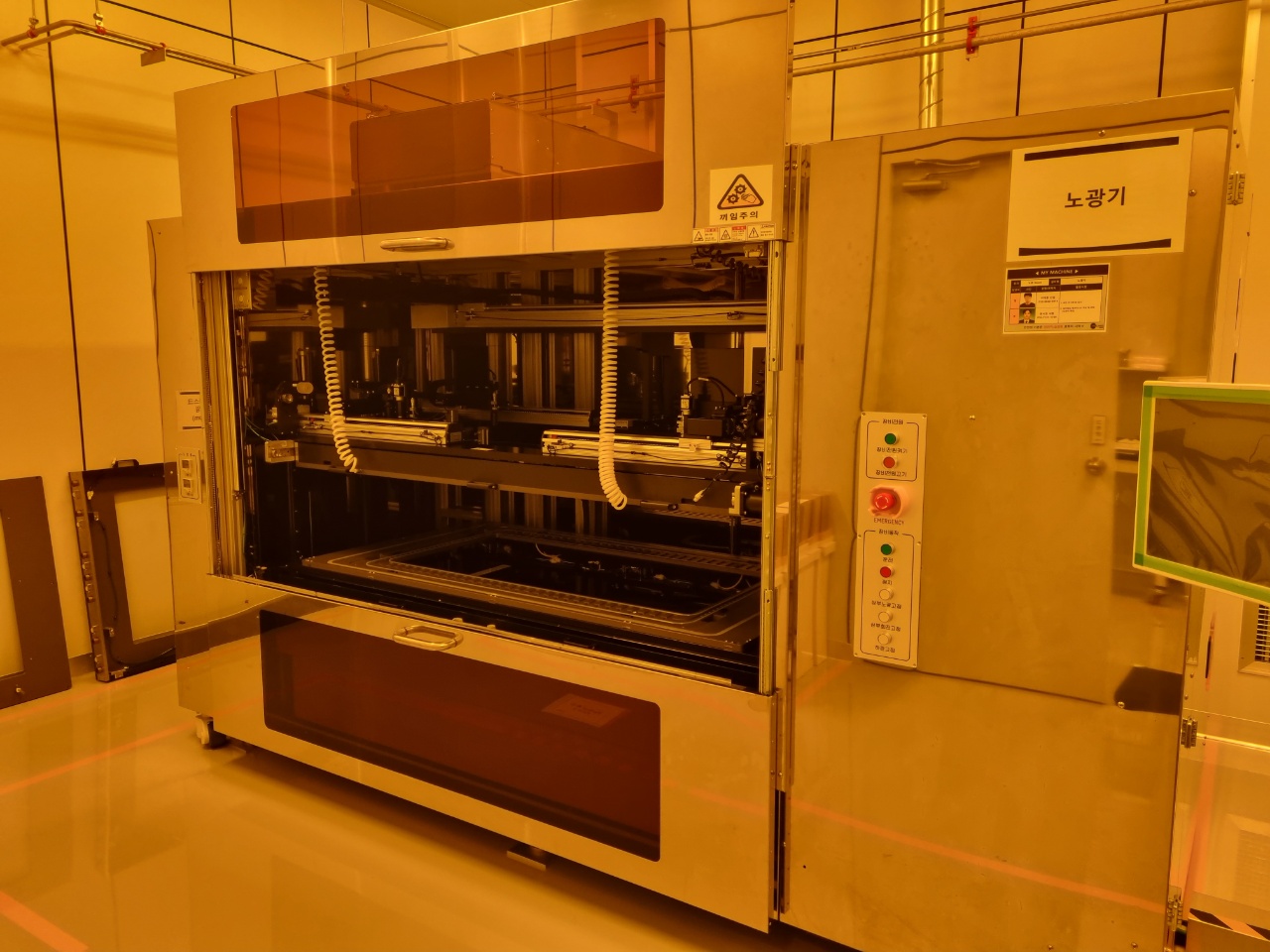}
  \caption[Large-sized bipolar UV exposure machine]{
  Large-sized bipolar UV exposure machine aligning top and bottom masks and irradiating UV through the masks.
  %Large-sized bipolar UV exposure machine that aligns top and bottom masks and irradiates UV light to dry film photoresist through the masks.
  }
  \label{fig:exposure}
\end{figure}

With the double-mask technique, the alignment of the top and bottom masks is crucial in producing GEM foils with the proper geometry. 
The allowable misalignment is under \SI{7}{\micro\meter}.
%The permissible misalignment is less than \SI{7}{\micro\meter}.
Fig.~\ref{fig:exposure} presents the large-sized bipolar 
% bipolar or bi-directional?
ultraviolet (UV) exposure machine that aligns the masks
through which UV light is irradiated onto dry film photoresist (DFR) laminated on both sides of the FCCL.
The misalignment was verified to be approximately \SI{5}{\micro\meter}.

%The size of the machine becomes the limiting factor of the size of GEM foils producible.
The machine's size becomes the limiting factor for the size of producible GEM foil.
The maximum area exposed to UV radiation and available area for pattern development is approximately 125$\times$58~\si{\centi\meter\squared}.
Emulsion glass masks are adopted for the alignment. 
%The GEM foil production by the consortium is more suitable for mass production than research and development (R\&D), owing to the substantially high price of emulsion glass masks.

Subsequent to pattern development on DFRs, Cu holes are generated through chemical wet etching, allowing for tunable hole diameters via adjusted etching conditions.
The process involves automated conveying machinery for DFR development and Cu etching.
%After the pattern is developed on DFRs, Cu holes are produced via chemical wet etching.
%The diameter of the holes is tunable by adjusting the etching conditions.
%The DFR development and Cu etching are carried out by automated conveying machinery.
PI holes are manually developed by immersing FCCLs in a bath filled with etchant.
Cu layers laminated on the PI layer function as masks for the PI etching processes.
A mixture of mono ethanolamine (MEA) and potassium hydroxide is used as the PI etchant\cite{Yoon:2023gyf}. 
%PI etching technique using MEA has been developed because it is less  toxic and less volatile whereas CERN MPT uses ethylene diamine (EDA).
%The acute lethal concentration (LC) for mice exposed to MEA vapor for \SI{2}{\hour} is larger than \SI{2430}{\milli\gram\per\cubic\meter}\cite{Pubchem_MEA}, on the other hand, the median LC (LC$_{50}$) of EDA vapor is \SI{300}{\milli\gram\per\cubic\meter}\cite{Pubchem_EDA}.
By adjusting the composition of the PI etchant, the geometry of developed PI holes is tunable.
To form the supporting areas, such as electrodes, a secondary Cu etching is carried out.
For this etching, emulsion glass masks are not necessary because the alignment of masks is not critical for the support areas. After the post processes, such as resistor soldering, the production is %completed.
complete.
\section{Quality validation}\label{sec::validation}
\begin{figure}[hbt]
  \centering
  \includegraphics[width=0.9\textwidth]{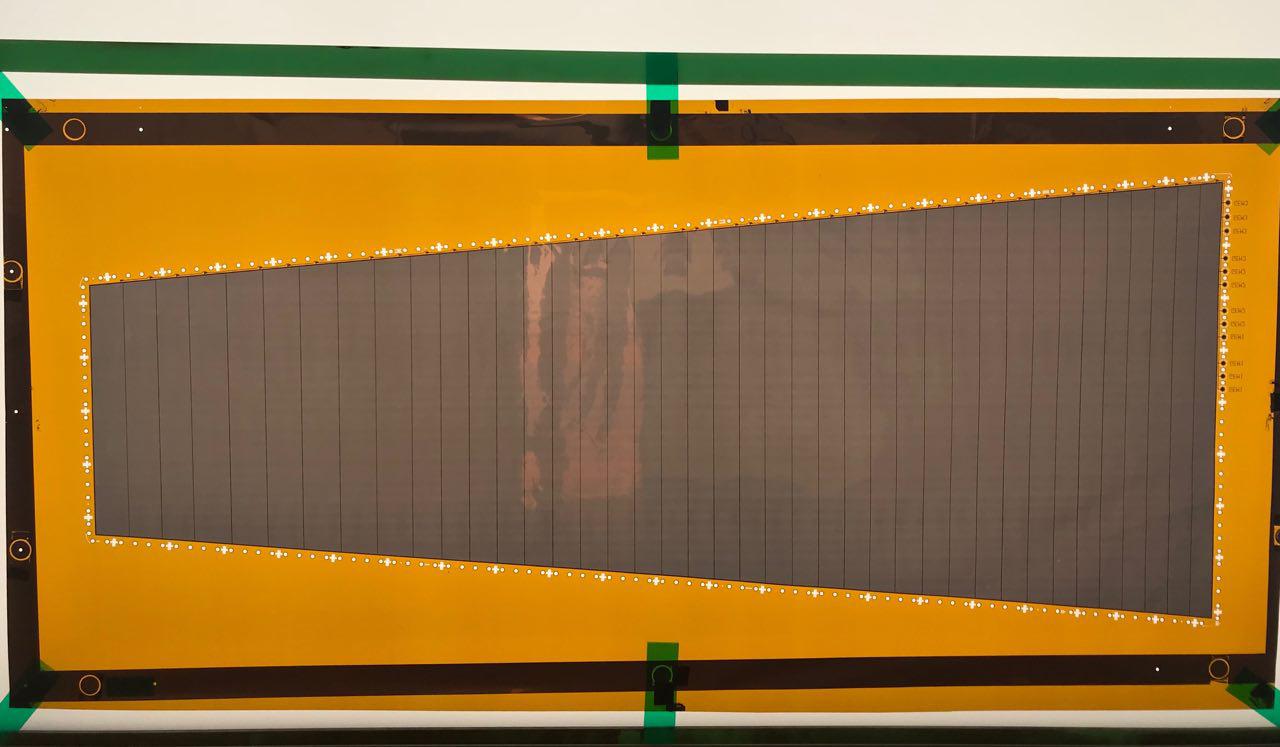}
  \caption[CMS GE1/1 short type GEM foil produced by the KCMS and Mecaro consortium]{
  CMS GE1/1 short-type GEM foil fabricated by the KCMS and Mecaro consortium for quality validation. 
  The active area has dimensions of \SI{106.1}{\centi\meter} in height, \SI{23.1}{\centi\meter} in short base and \SI{42.0}{\centi\meter} in long base.}
  \label{fig:Mecaro_GE11_Foil}
\end{figure}

%After the consortium succeeded in fabricating the CMS GE1/1 short %type GEM foils they were validated.
%
After fabrication, the GE1/1 short-type GEM foils were validated.
Fig.~\ref{fig:Mecaro_GE11_Foil} shows a GE1/1 foil used for the validation.
The GE1/1 foils are single-segmented foils: only the Cu layer facing the drift electrode (drift side) is segmented to control the amount of energy released when a discharge occurs. The opposite layer, facing the readout electrodes (readout side), is not segmented.
The validation process utilized the CMS GEM detector assembly facility and CERN's gamma irradiation facility (GIF++) \cite{PFEIFFER201791}.
%The GEM detector assembly facility and the CERN gamma irradiation facility++ (GIF++) \cite{PFEIFFER201791} were utilized for the validation.

\subsection{Optical inspection}
To begin, optical inspections were conducted to assess the quality of hole development.
Given that the properties of GEM foils rely heavily on geometry, it is crucial to ensure that the designed geometry is accurately realized.

\begin{figure}[hbt]
  \centering
  \subfloat[Drift side]{
    \includegraphics[width=0.3\textwidth]{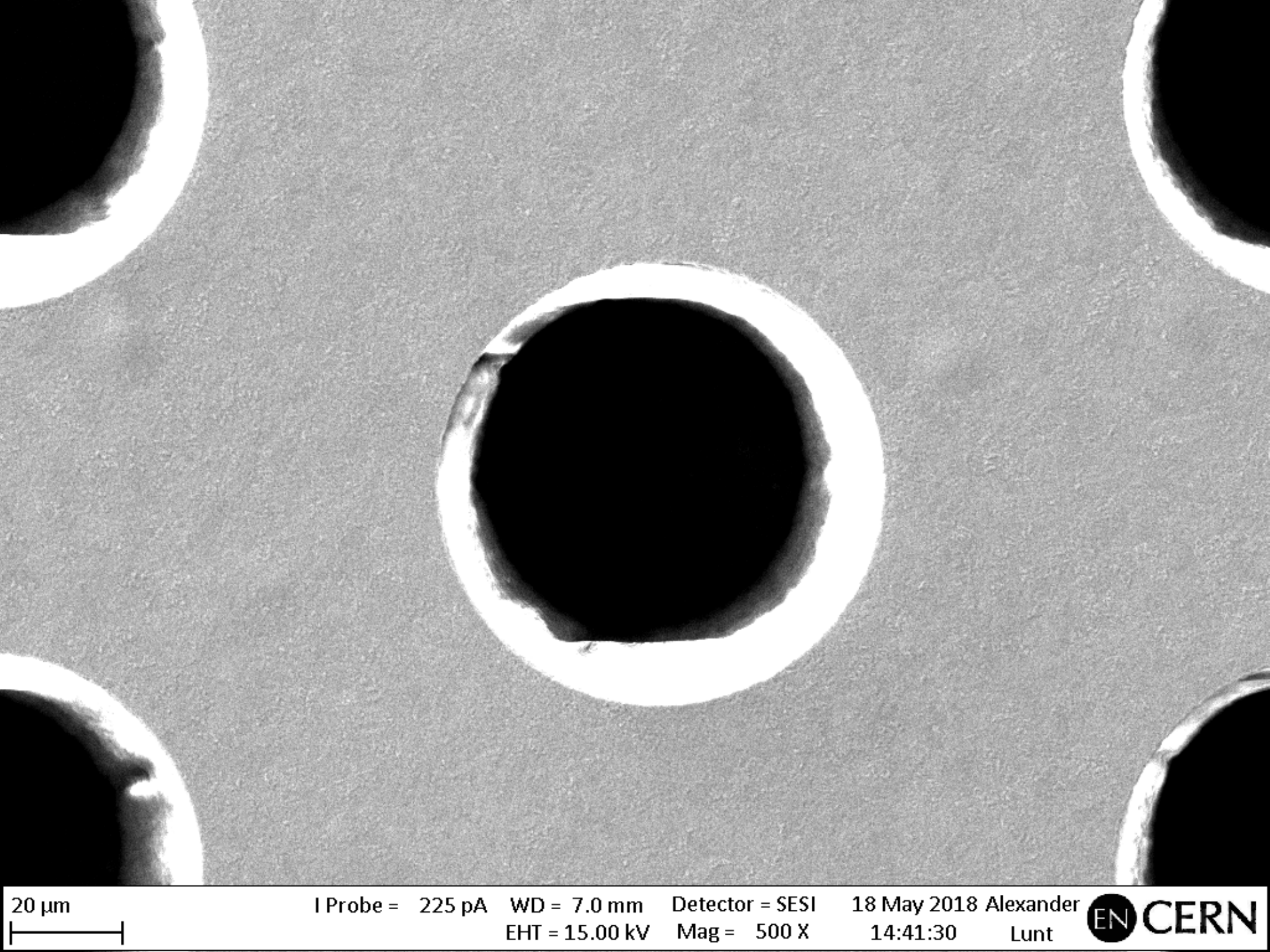}
  }
  \subfloat[Readout side]{
    \includegraphics[width=0.3\textwidth]{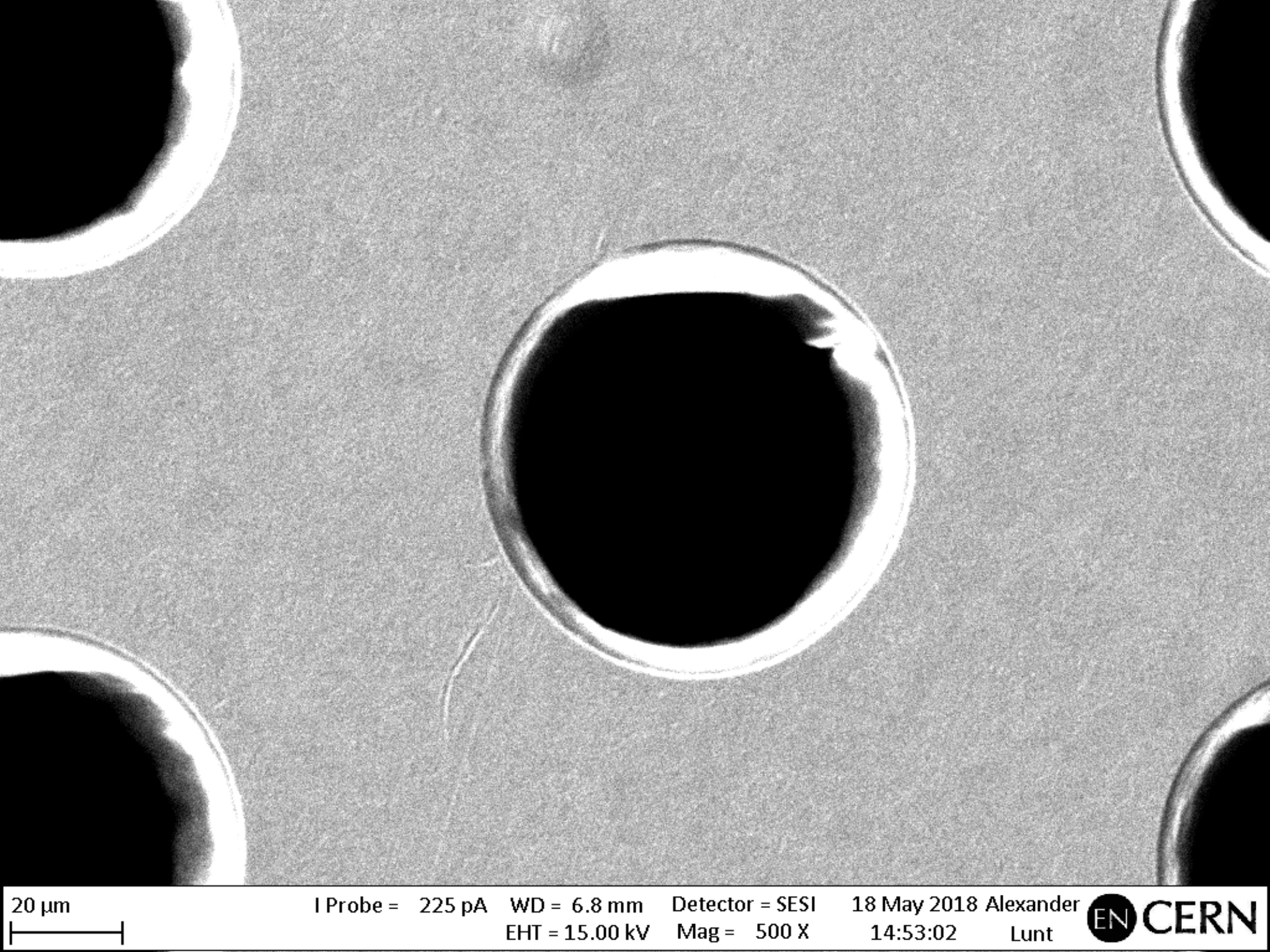}
  }
  \subfloat[Cross section]{
    \includegraphics[width=0.3\textwidth]{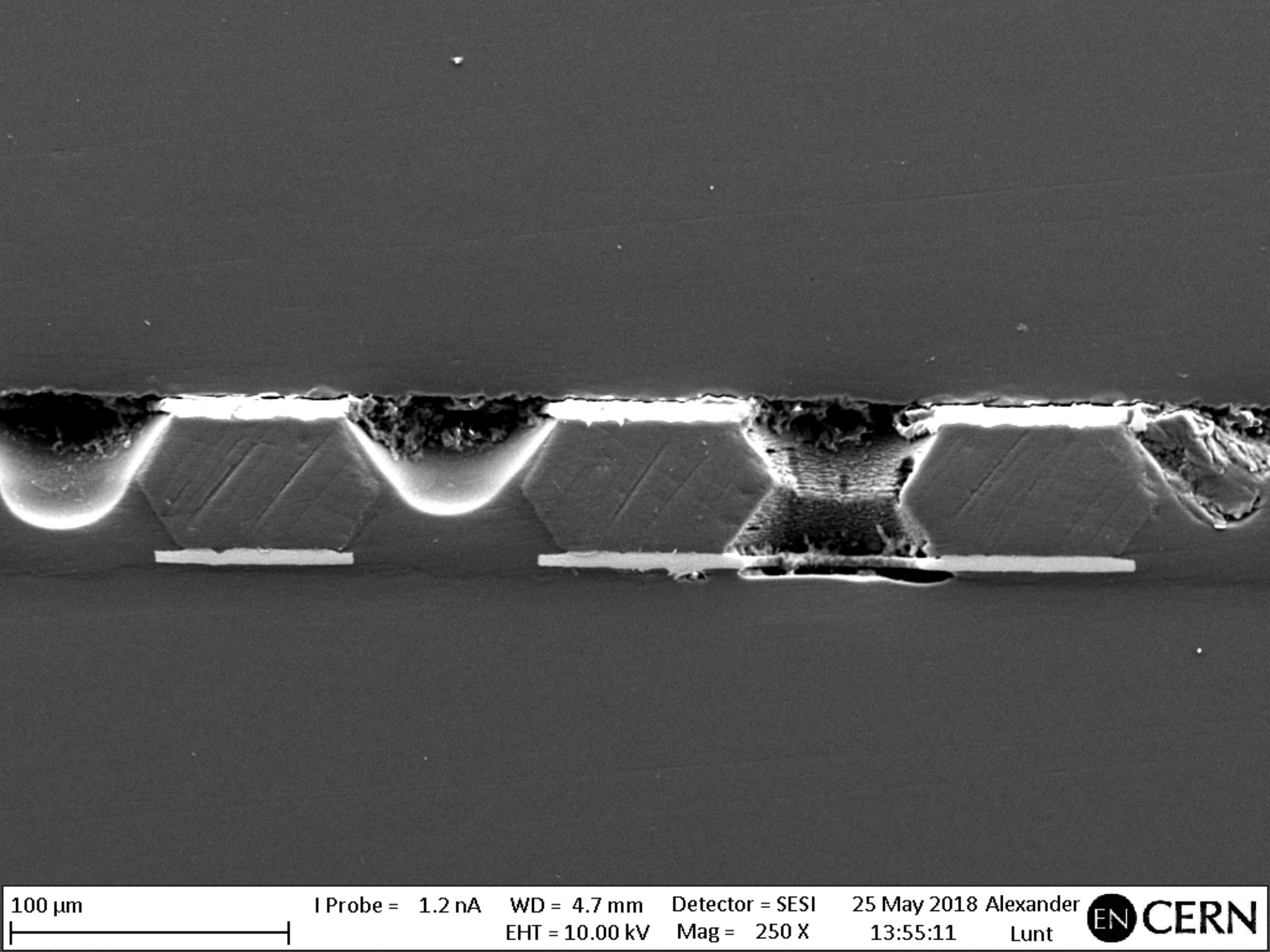}
  }
  \caption[SEM images of KCMS and Mecaro GEM foil]{
  SEM images of a GEM foil fabricated by Mecaro. 
  In the cross-section view (c), the material visible inside the holes is resin that was added to protect the foil during the cutting and polishing process.
  }\label{fig:sem}
\end{figure}

Scanning electron microscope (SEM) images were taken to evaluate the quality of the hole development.
Fig.~\ref{fig:sem} shows SEM images of the drift and readout sides and cross-section of a fabricated GEM foil. 

%The diameters of the holes were measured with a manually %controlled optical microscope, which was well calibrated with %circular patterns of precise diameters.
%For 15 positions on the GEM foil, 30 holes for each position, the %diameters of 450 holes were measured.
%The measured Cu (PI) diameters were \num{70.67 \pm 2.04} %(\num{50.69 \pm 2.04})\,\si{\micro\meter}.
%No significant difference in the diameters was observed based on %position; however, statistically precise measurement were not %possible with the manually controlled microscope.

The diameters of the holes were first measured with a manually controlled optical microscope that was calibrated with circular patterns of accurate diameters.
For 15 positions on the GEM foil and 30 holes for each position, the diameters of 450 holes were measured.
The Cu (PI) diameters were \num{70.7 \pm 2.0} (\num{50.7 \pm 2.0})\,\si{\micro\meter}.
No significant difference in the diameters was observed based on position. 

\begin{figure}[hbt]
  \centering
  \subfloat[Cu hole, drift side]{
    \includegraphics[width=0.5\textwidth]{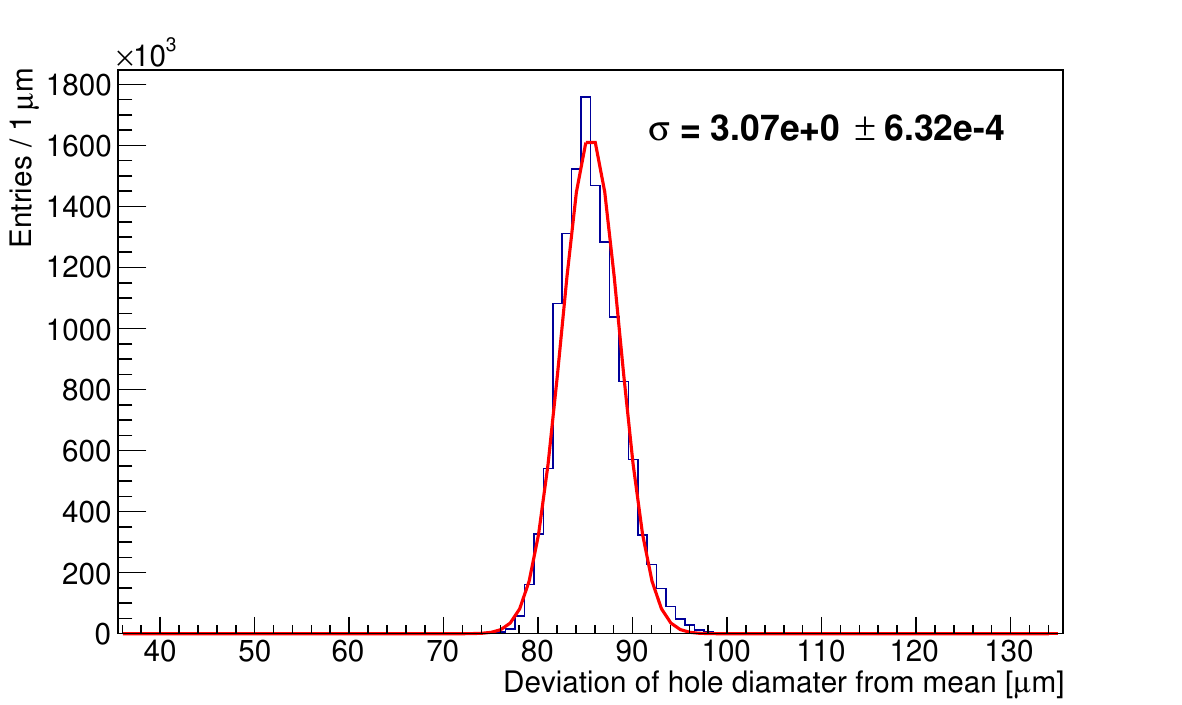}
  }
  \subfloat[Cu hole, readout side]{
    \includegraphics[width=0.5\textwidth]{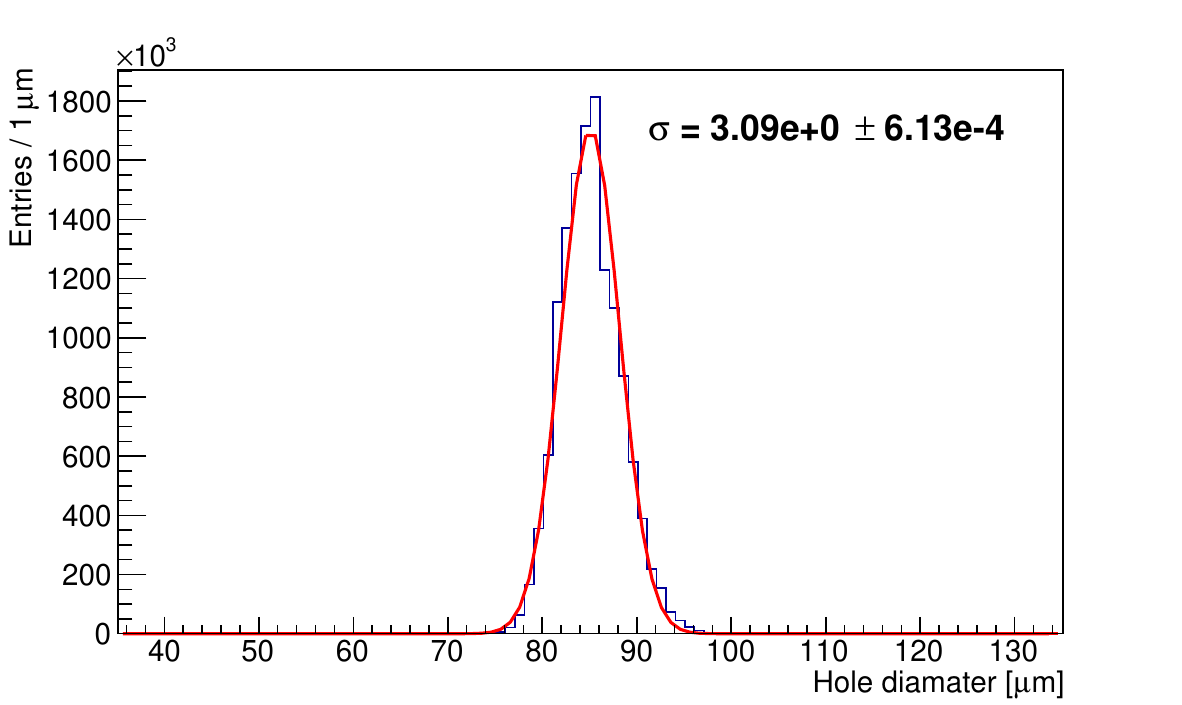}
  }\\
  \subfloat[PI hole, drift side]{
    \includegraphics[width=0.5\textwidth]{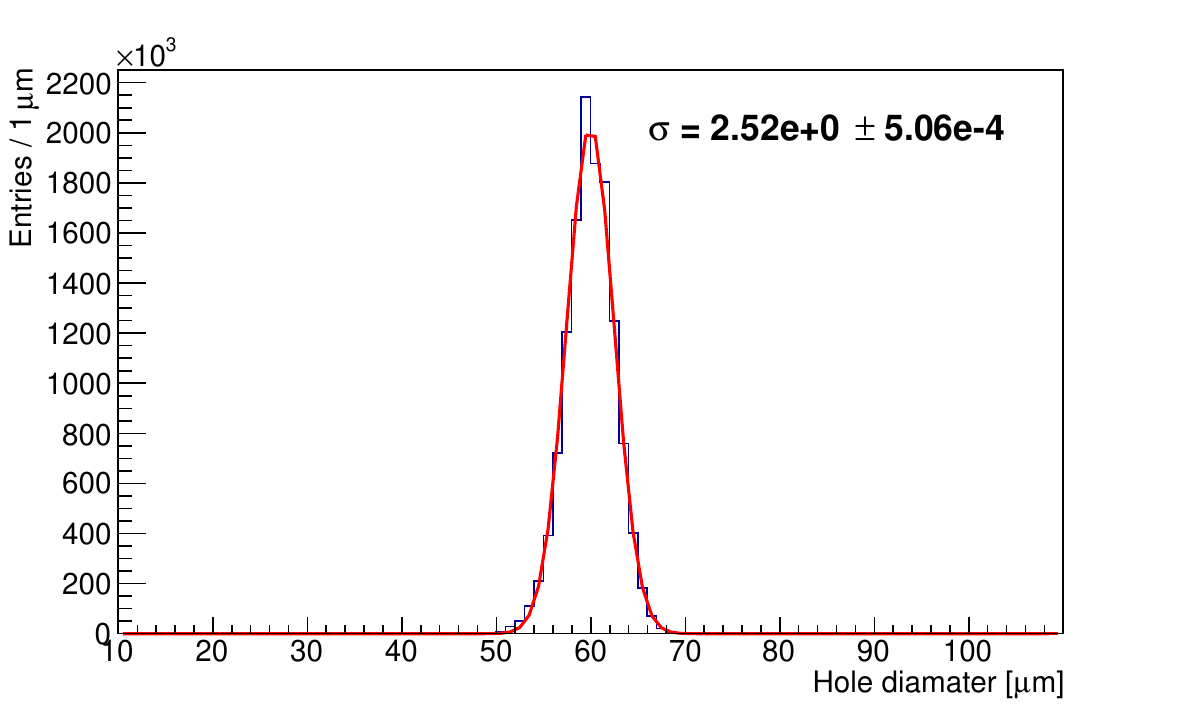}
  }
  \subfloat[PI hole, readout side]{
    \includegraphics[width=0.5\textwidth]{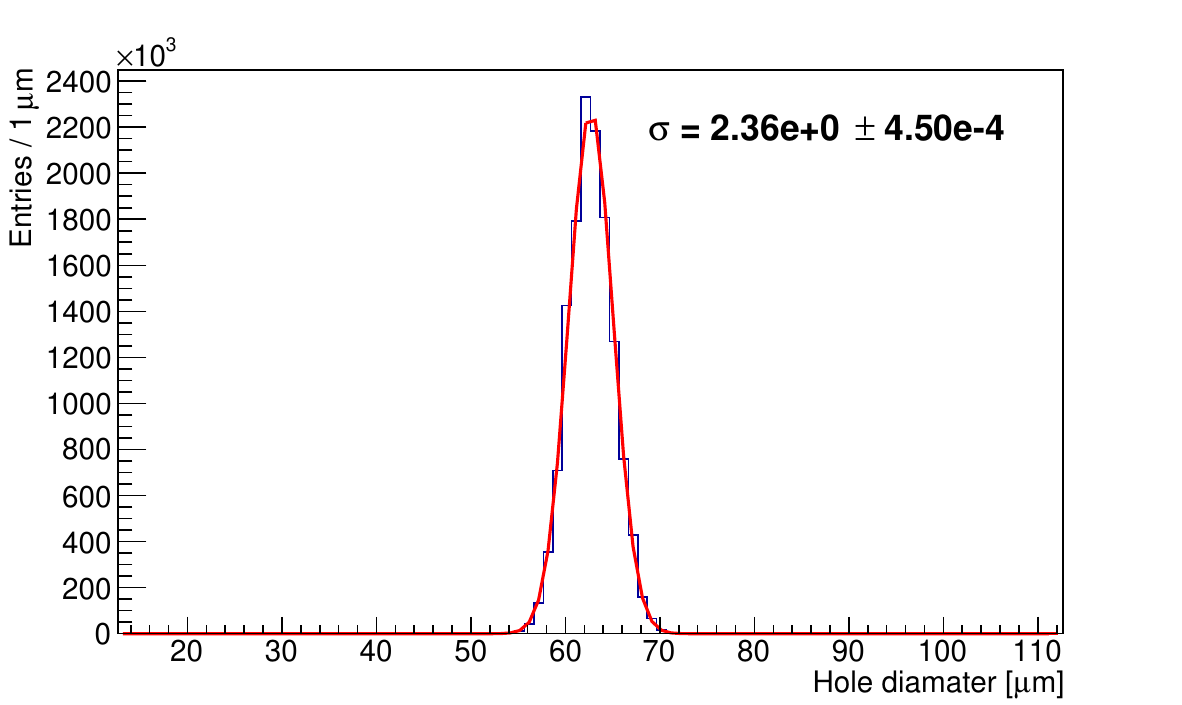}
  }
    \caption[Distributions of hole diameters]{
    Distributions of hole diameters.
    The first (second) row presents the deviation of Cu (polyimide) hole diameters, while the first (second) column illustrates the deviation measured from the drift (readout) side of the GEM foil. \cite{Posik_private}.
    }\label{fig:hole_diameter_distribution}
\end{figure}

Since hole diameter depending on position within a foil is an important factor in uniformity of detector response, an automated 2-D CCD camera scanner \cite{POSIK201510} is employed to measure the diameter of every hole in a foil.
The distribution of hole diameter is shown in Fig.~\ref{fig:hole_diameter_distribution}.
The distribution is well described by a single Gaussian with variance of the Cu (PI) diameter of \num{3.08} (\num{2.49})\,\si{\micro\meter}. 
The difference in central values between the manual microscope and the scanner measurements is attributed to the scanner introducing a bias during hole recognition and diameter measurement.
%The scan results did not reveal any bias in hole diameter with respect to position within a foil \cite{Posik_private}.

%No significant difference in the variance of hole diameters was observed between the drift side and the readout side, showing the successful application of the double-mask technique.

However, it was observed that the diameters of PI holes varied by production batches.
Review of the PI etching process pointed to the denaturing of the etchant under reaction with carbon dioxide in air, which was not considered previously. The process was modified so that the etchant is tuned before use with each PI batch.
In addition, a strict quality control (QC) protocol is applied.
The diameters of Cu holes are observed to be robust.

\subsection{Long-term measurement of leakage current and applied voltage in dry nitrogen}
%\subsection{Long-term leakage current measurement at dry condition}
After optical inspection of a foil, its leakage current was measured to assess its cleanliness.
%After the optical inspection, a leakage current of the foil was measured to assess the cleanliness of the foil.
Because contaminants on the foil can provide electrical bridges between the Cu layers, the foil's impedance is a measure of its cleanliness.
%
%The impedance of the foil  depends significantly on the quality of the foil's cleanliness, as contaminants on the foil function as additional electrical bridges between the Cu layers.
%
It is also important to identify defects in the structure of the foil that may induce sparks. In principle this could be accomplished by optical inspection. However, owing to the microscopic nature of the foil, optical inspection is impractical because it takes a long time and can even be destructive in some cases. Instead, defects can be identified during leakage current measurement by trips of the power supply due to an abrupt increase in current.
%
%It is also important to identify defects in the structure of the foil can be identified by electrical assessment because they may induce sparks.
%
%Owing to the microscopic nature of the GEM foil, the optical inspection is usually impractical as it takes a protracted amount of time and can even be destructive in some cases.
%Hence, the leakage current measurement are a crucial and practical tool in evaluating the quality of GEM foils.
%
Thus, measurement of leakage current and trips in applied voltage are practical tools to evaluate the cleanliness and quality of GEM foils.

%A leakage current was measured in accordance with the standard QC methodology invented for the CMS upgrade \cite{Abbaneo_2015}.
%
Leakage current and frequency of discharges were measured with high voltage applied to the GEM foils using a QC methodology developed for the CMS GEM upgrade \cite{Abbaneo_2015}.
%
%A leakage current and the frequency of discharges were measured for the QC procedures with high voltage applied to GEM foils.
%The QC procedures consist of two parts, the named QC-fast and QC-long. 
%
The QC procedure consists of two parts, called QC-fast and QC-long.
%The primary difference between the QC-fast and QC-long are the durations of these QCs and the atmosphere in which the QCs performed.
The differences between QC-fast and QC-long are the time duration and atmospheric environment. 

The QC-fast is a short test and is performed in ordinary air.
Preliminary assessment of a foil is obtained via the QC-fast. The QC-long lasts for at least \SI{7}{\hour}, depending on the cleanliness quality of the GEM foils.
The QC-long is performed inside a plexiglass container filled with dry nitrogen to control relative humidity.

\begin{figure}[hbt]
  \centering
  \includegraphics[width=0.7\columnwidth]{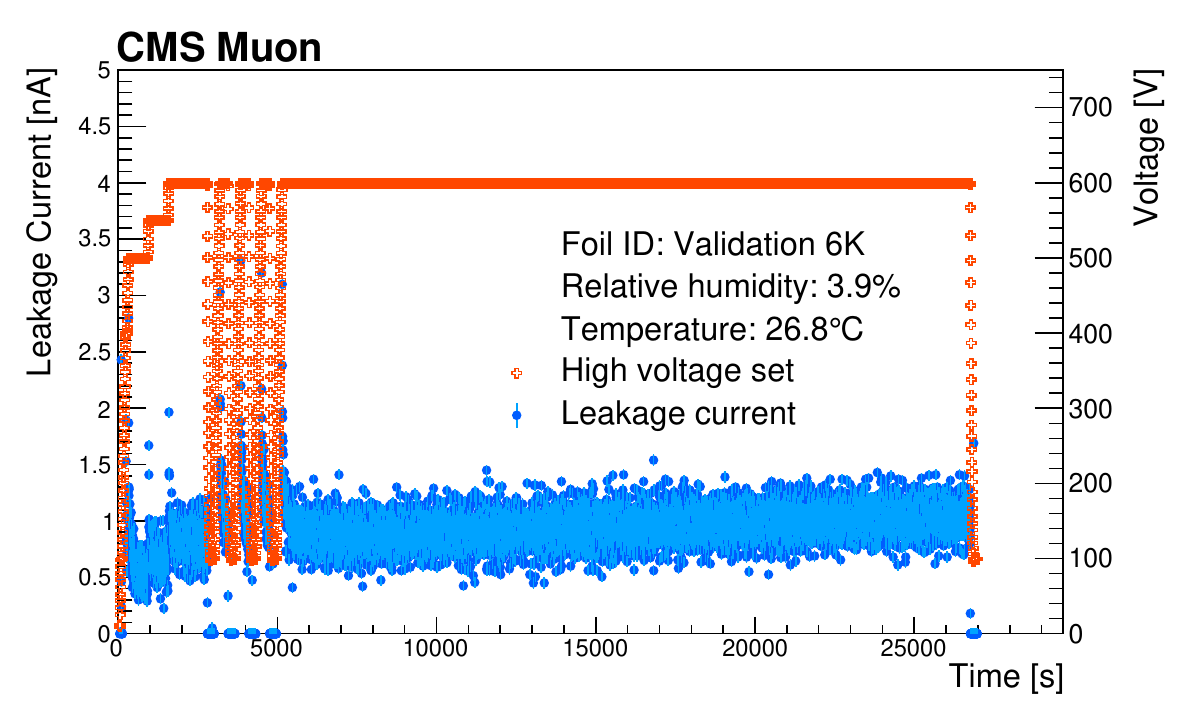}
  \caption[Result of QC-long]{Conventional example of QC-long test result of GEM foil fabricated by KCMS and Mecaro. The horizontal axis represents duration of the test. Applied high voltage to the GEM foil is shown by the red markers. Blue markers denote leakage current. The rapid changes in applied voltage at the beginning of the test are intended to stress the foils.}
%\caption[Result of QC-long]{Conventional example of QC-long test result. The horizontal axis represents duration of the test. Applied high voltage to GEM foil is shown by the red markers. Blue markers denote leakage current. The rapid changes in applied voltage at the beginning of the test are intended to give stress on foils.}
  \label{fig:qc_long}
\end{figure}

Fig.~\ref{fig:qc_long} presents an example of a QC-long result showing
stable, low leakage current without discharge which can be identified by power supply trips.
Only foils that pass the QC-long test are used in assembling detectors.
If a foil shows leakage current more than \SI{5}{\nano\ampere} or discharges that occur too frequently, the foil is repeatedly cleaned chemically and physically.

%Fig.~\ref{fig:qc_long} presents an example of the QC-long result.
%Stable and low leakage current without discharge was observed for the corresponding foil.
%Discharges are identified by trips of a power supply owing to abrupt increase in current.
%Only foils that pass the QC-long test are adopted in assembling the detectors.
%If the leakage current is of over \SI{5}{\nano\ampere} or discharges occur too frequently, the foil should be repeatedly cleaned chemically and physically.

%Figure 5 should be made larger.

\subsection{Quality validation with CMS GE1/1 short-type detectors}\label{subsec::detector_validation}
\begin{figure}[hbt]
  \centering
    \includegraphics[width=0.7\textwidth]{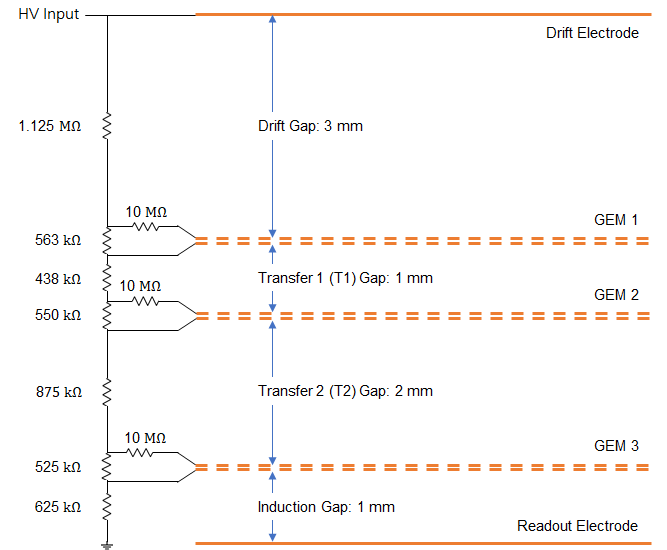}
  \caption[High voltage supply scheme]{High voltage supply scheme of the CMS GE1/1 detectors. High voltage is distributed to drift electrodes and GEM foils with a voltage divider. Surface mounted protection resistors of \SI{10}{\mega\ohm} are positioned on the drift side of GEM foils, serving as the path for applying high voltage to the active regions of the GEM foils.}
  \label{fig:hv_supply}
\end{figure}

Korean GEM foils that passed the QC-long test were used to assemble four CMS GE1/1 short-type detectors. 
The GE1/1 detectors are the triple-GEM detectors with 3/1/2/1 \si{\milli\meter} gap configuration and operate using a mixture of \SI{70}{\percent} Ar and \SI{30}{\percent} of CO$_{2}$.
High voltage is supplied using an in-house designed voltage divider as described by Fig.\ref{fig:hv_supply}.
%The divider consists of series of resistors with \SI{1.125}{\mega\ohm}, \SI{560}{\kilo\ohm}, \SI{438}{\kilo\ohm}, \SI{550}{\kilo\ohm}, \SI{875}{\kilo\ohm}, \SI{525}{\kilo\ohm} and \SI{625}{\kilo\ohm} resistance to supply high voltage to the drift gap, . 
The design of the GE1/1 detectors is described in detail in \cite{Colaleo:2021453}. 
The properties of the detectors were measured to validate the quality of the Korean GEM foils.
Upon assembly completion, integrity verification was conducted through gas tightness and high voltage circuit tests, ensuring the assembly's soundness before proceeding to measure other detector properties, as outlined in \cite{VENDITTI2018}.
%After assembly, gas tightness and high voltage tests were performed to verify the integrity of the assembly before measurement of key chamber properties, following the procedures described in \cite{VENDITTI2018}. 

%% due to comments from journal referee
%Korean GEM foils that passed the QC-long test were used to assemble four CMS GE1/1 short-type detectors. 
%The design of the GE1/1 detectors, which operate with an Ar/CO$_{2}$ (\SI{70}{\percent}/\SI{30}{\percent}) mixture, is %described in detail in \cite{Colaleo:2021453}. 
%High voltage is supplied using an in-house designed voltage divider.
%The divider consists of series of resistors with \SI{1.125}{\mega\ohm}, \SI{560}{\kilo\ohm}. 
%The properties of the detectors were measured to validate the quality of the Korean GEM foils.
%After assembly, gas tightness and high voltage tests were performed to verify the integrity of the assembly before measurement of key chamber properties, following the procedures described in \cite{VENDITTI2018}. 

%Using Korean GEM foils, which passed the QC-long test, CMS GE1/1 short-type detectors were assembled.
%CMS GE1/1 detectors are triple-GEM detectors filled with an Ar/CO$_{2}$ (\SI{70}{\percent}/\SI{30}{\percent}) mixture.
%High voltage is supplied to the detector elements using an in-house designed voltage divider.
%Further details can be found in \cite{Colaleo:2021453}.
%The properties of the detectors were measured to validate the quality of the Korean GEM foils.
%Subsequently, four detectors were assembled. After the assembly, gas tightness and high voltage circuit tests were performed to verify the integrity of the assembly before measurements of the other properties of detectors were carried out, as described in \cite{VENDITTI2018}.

\textbf{Effective Gain}: 
%The first measured item is the effective gas gain. The effective gain of detectors was calculated by measuring the readout current ($I_{RO}$) and event rate ($R$) while the detectors were exposed to x-ray from Amptek Mini-X2 generator with an Ag transmission target.
%
Effective gain ($G$) is a basic property of a GEM detector and can be used to compare detectors assembled with Mecaro foils with those assembled with CERN foils.
The effective gain was calculated from the readout current ($I_{RO}$) and event rate ($R$) of detectors irradiated with x-rays from an Amptek Mini-X2 generator with an Ag transmission target.
%
%$N_{p}$ is estimated as \num{346 \pm 2.9} by comparing responses of detector to x-ray from the generator and x-ray from $^{55}$Fe.
%Further details can be found in \cite{VENDITTI2018}
%
$I_{RO}$ was measured using a Keithley 6487 picoammeter.
%The effective gain was calculated as $G=\frac{I_{RO}}{R\times{}q_{el}\times{}N_{p}}$, where $q_{el}$ is the electron charge and $N_{p}$ is the number of primary electrons produced by the photo-electron originating from fluorescence of a copper atom in the drift electrode that has absorbed an x-ray from the generator.
The calculation of effective gains followed the formula $G=\frac{I_{RO}}{R\times{}q_{el}\times{}N_{p}}$, where $q_{el}$ is the electron charge and $N_{p}$ signifies the number of primary electrons generated through ionization due to the fluorescence of a Cu atom within the drift electrode, which has absorbed an x-ray from the generator.
$N_{p}$ is estimated as \num{346 \pm 2.9} by comparing the response of a detector to x-rays from the generator with the response to x-rays from $^{55}$Fe,
as described in \cite{VENDITTI2018}.

%Fig.~\ref{fig:gas_gain} presents the results of the effective gain measurements.
%At the divider current of \SI{660}{\micro\ampere}, the defined reference working current, the effective gains were measured as \SIrange[tophrase={--}]{7.0E3}{1.6E4}{}.
%The obtained results are consistent with the gains of detectors with CERN foils as presented in \cite{VENDITTI2018}.

\begin{figure}[hbt]
    \centering
    \includegraphics[width=0.7\textwidth]{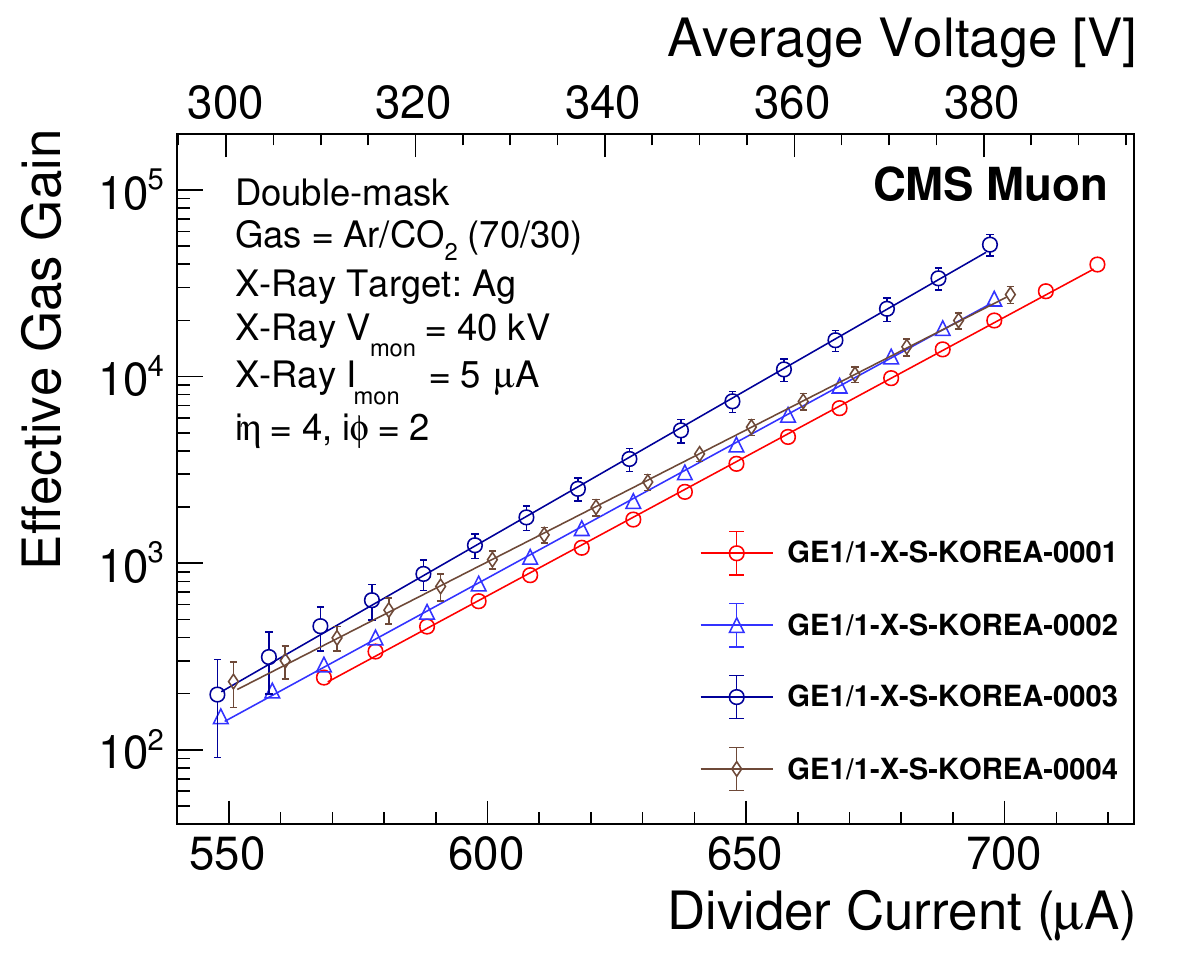}
    \caption[Effect gas gain curves]{Effective gain versus both of HV divider current and the average voltage applied to GEM foils for four CMS GE1/1 short-type detectors assembled with Mecaro foils.}
    \label{fig:gas_gain}
\end{figure}

Measurements of effective gain versus both of high voltage divider current and the average voltage applied to GEM foils for four detectors are presented in Fig.~\ref{fig:gas_gain}.
A current of \SI{660}{\micro\ampere} provides a reference working point that can be used to compare the effective gas gain of different detectors.
%For the four detectors with Mecaro foils, the effective gains were measured to be in the range  \SIrange[tophrase={--}]{5.3E3}{1.2E4}{}, consistent with the gains of detectors with CERN foils \cite{VENDITTI2018}.
For the four detectors with Mecaro foils, the effective gains were measured to be in the range  \SIrange{5.3E3}{1.2E4}{}, consistent with the gains of detectors with CERN foils \cite{VENDITTI2018}.

\textbf{Response uniformity}: 
It is important that the signal response from strips in all regions of the detector is uniform to ensure good trigger and track reconstruction efficiencies over the geometrical acceptance of the detector.
The readout strips of the CMS GE1/1 detector are comprised of a total of 3072 strips, organized in a matrix of 8 rows in the rapidity direction and 384 columns in $\phi$ direction. 
To measure response uniformity, as elaborated in detail in \cite{VENDITTI2018}, adjacent sets of four readout strips in the $\phi$ columns are grouped together, segmenting the detector into a total of 768 virtual sectors.
For each sector, the spectrum of charge was obtained while the detectors were exposed to x-rays. 
The detector strips were read out by APV25 front-end ASICs \cite{FRENCH2001359} and ADCs.
The ADC spectrum for each sector was fit with a Cauchy distribution function and a fifth order polynomial to determine the ADC value of the Cu fluorescence peak. 

%The distribution of sector peak ADC values for each of the four detectors with Mecaro foils is shown in Fig.~\ref{fig:gain_uniformity}.
%The distributions were fit with a Gaussian function to obtain mean ($\mu$) and variance ($\sigma$).
%Response uniformity, $\sigma/\mu$, of the four detectors are in the range \SIrange{10.2}{16.2}{\percent}, consistent with the uniformities of detectors with CERN foils \cite{VENDITTI2018}

\begin{figure}[hbt]
  \centering
  \subfloat{
    \includegraphics[width=0.35\columnwidth]{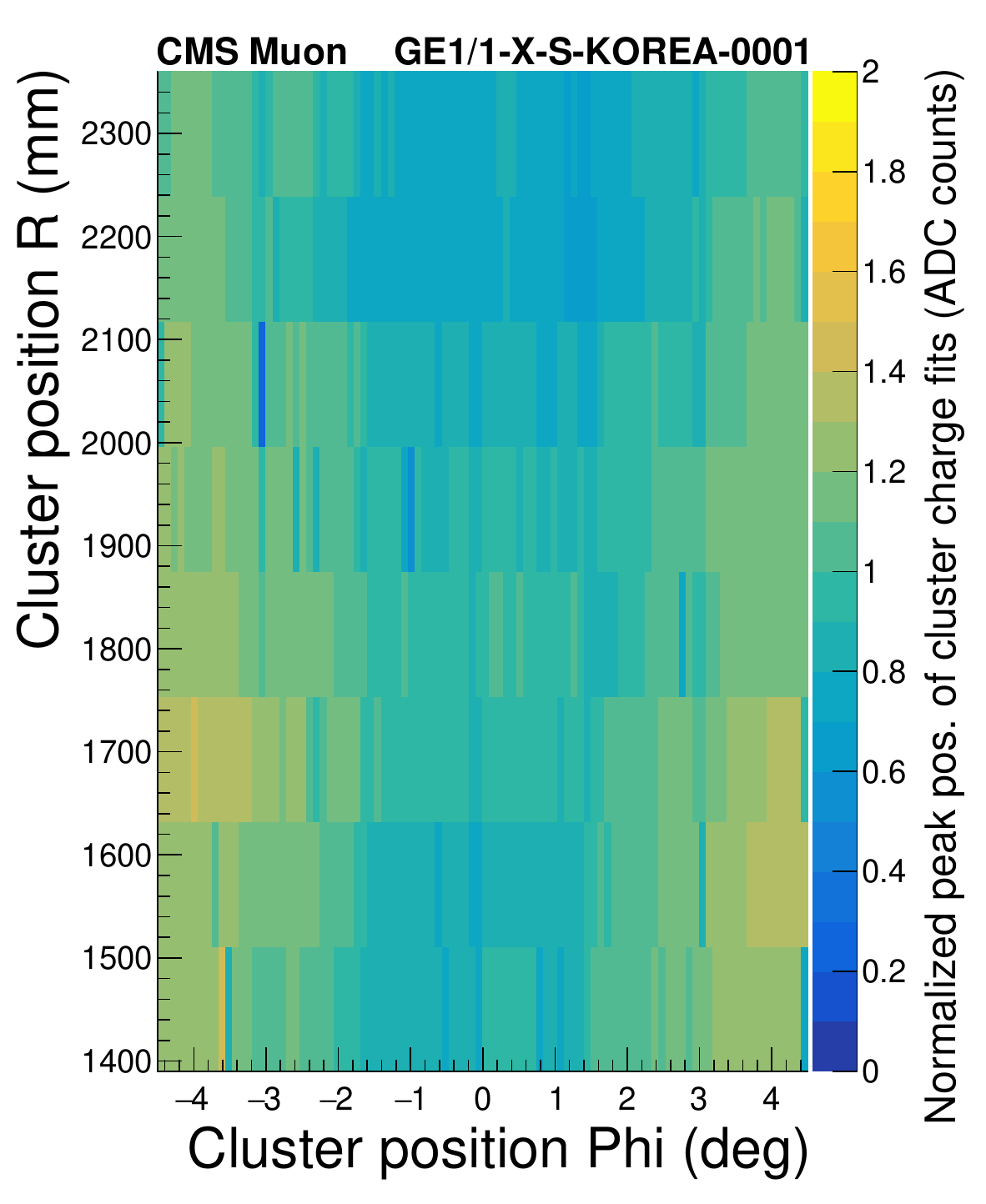}
  }
  \subfloat{
    \includegraphics[width=0.35\columnwidth]{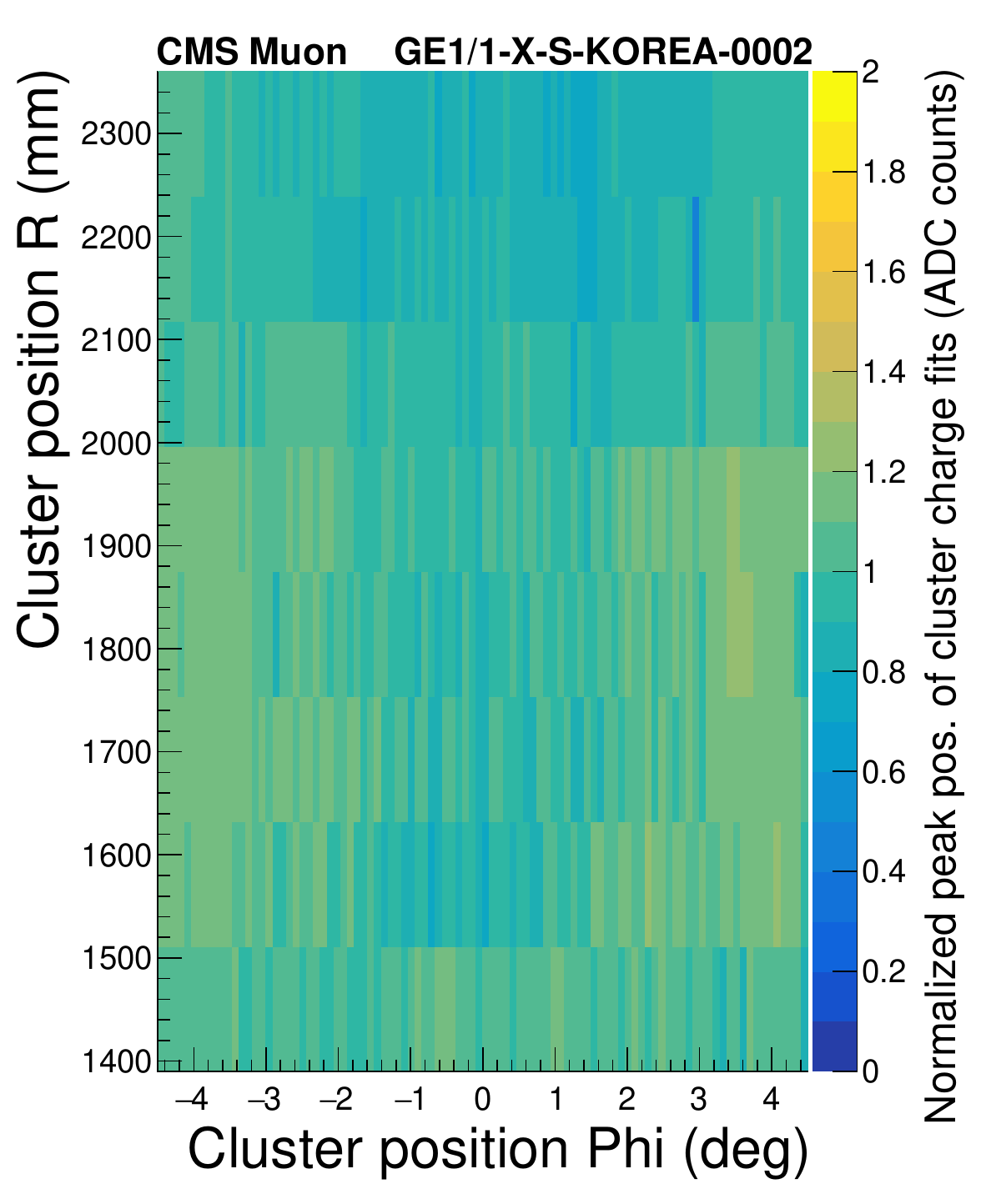}
  }\\
  \subfloat{
    \includegraphics[width=0.35\columnwidth]{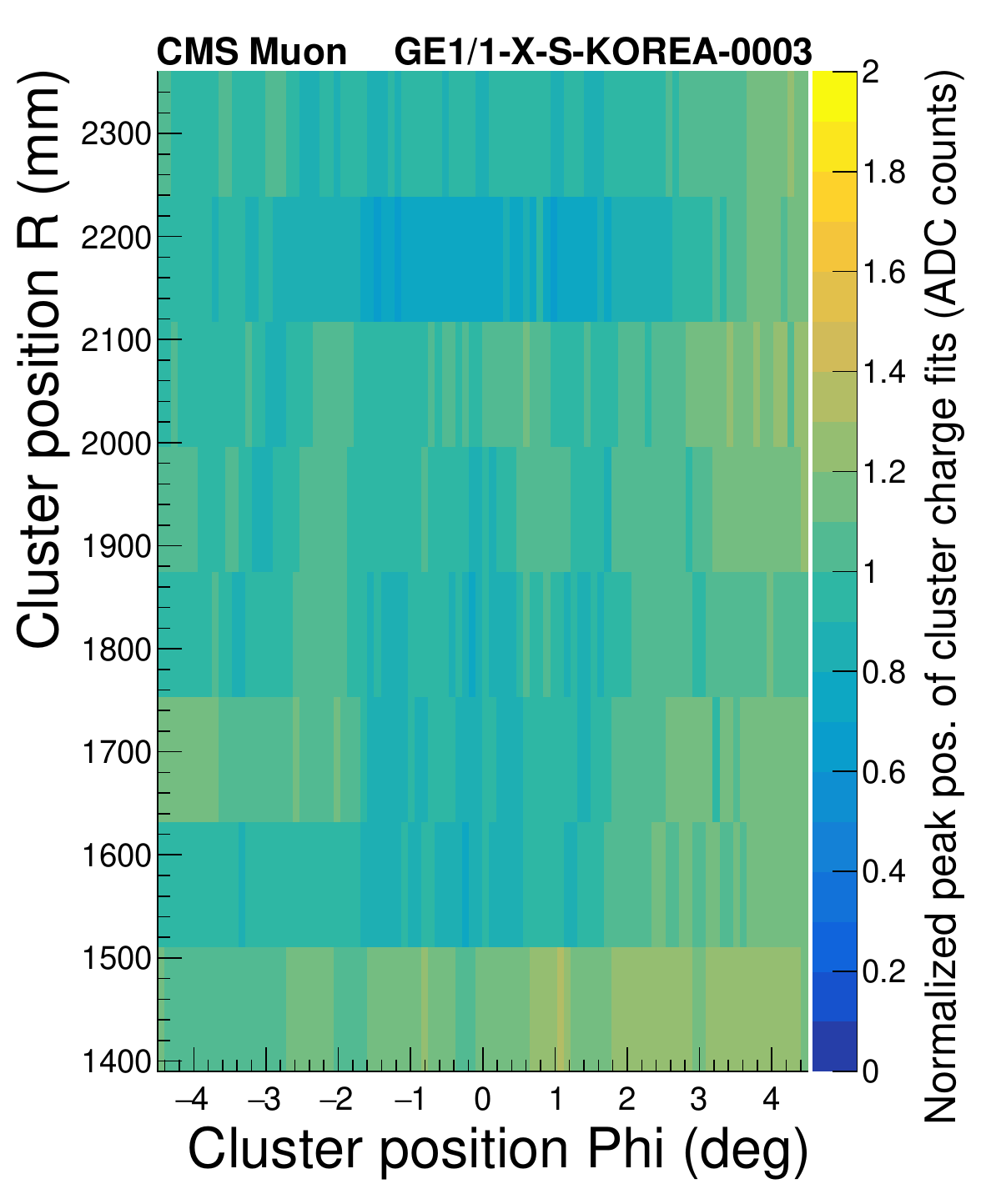}
  }
  \subfloat{
    \includegraphics[width=0.35\columnwidth]{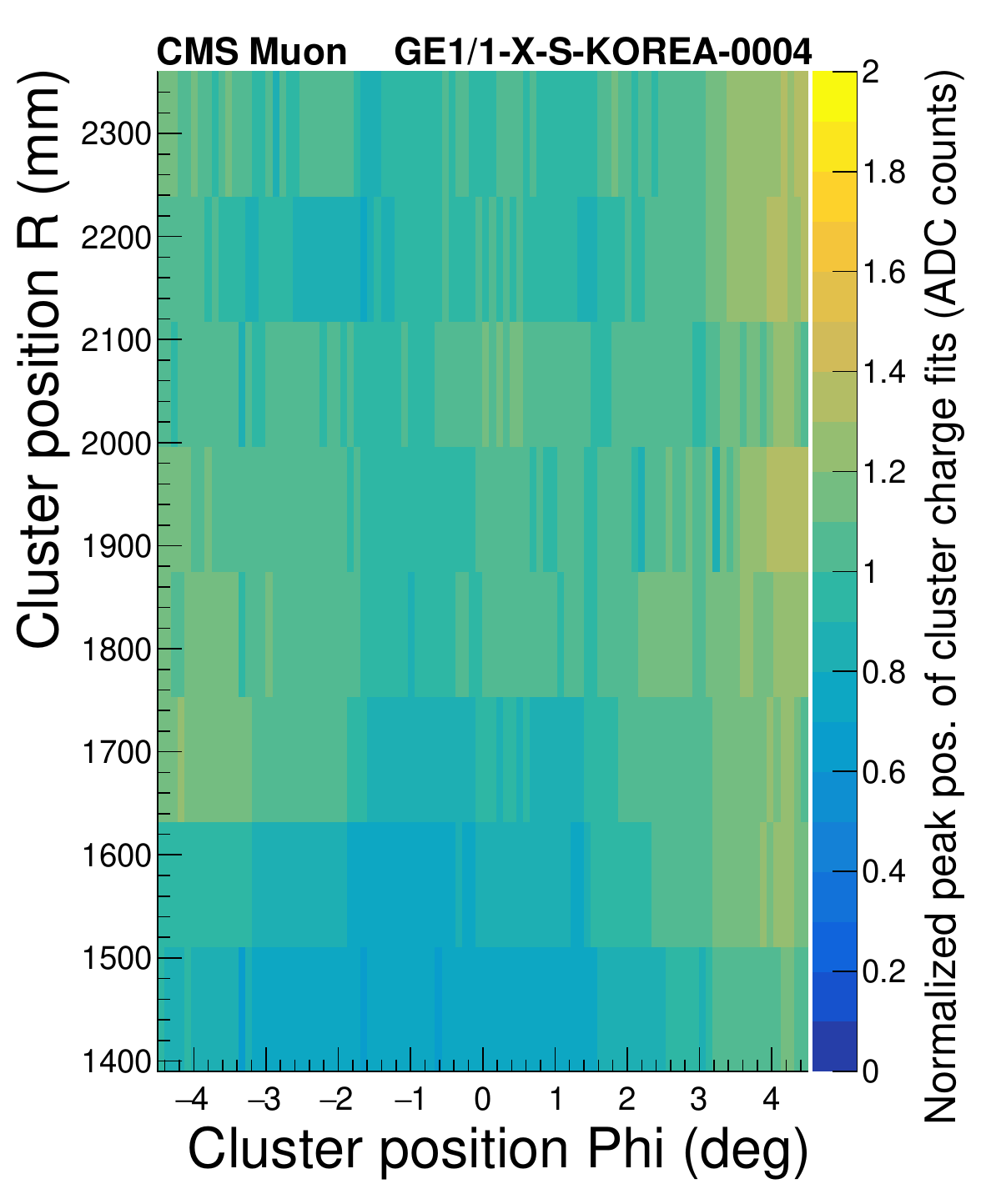}
  }
  \caption[Response uniformity 2D]{
  Gain variation across a GE1/1 chamber with Mecaro foils normalized to the chamber average. 
  Note that the coordinates are plotted in angular distance from the centerline of the chamber and radial distance from the beamline in the CMS experiment.
    }\label{fig:2d_gain_uniformity}
\end{figure}

\begin{figure}[hbt]
  \centering
  \subfloat{
    \includegraphics[width=0.45\columnwidth]{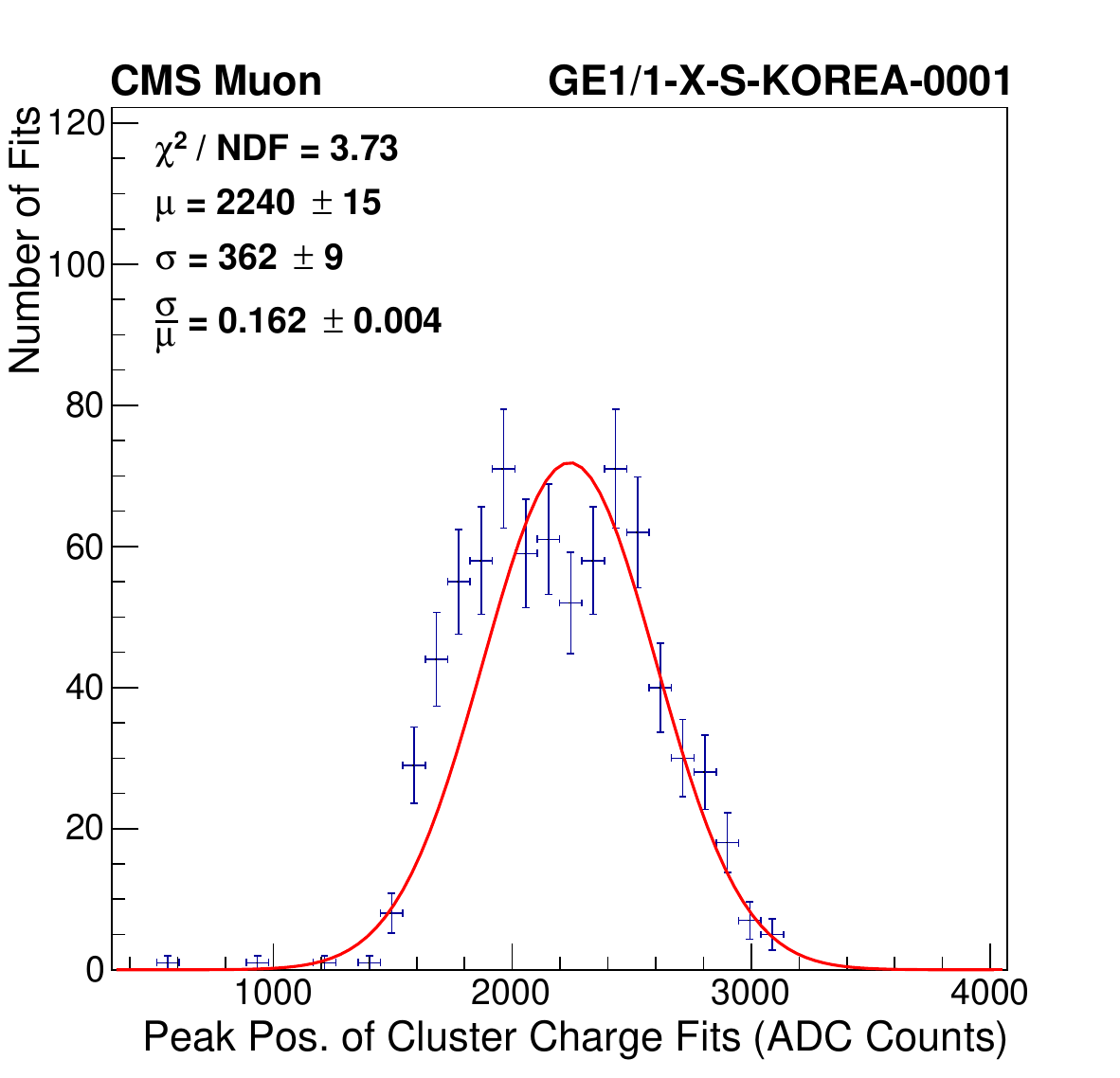}
  }
  \subfloat{
    \includegraphics[width=0.45\columnwidth]{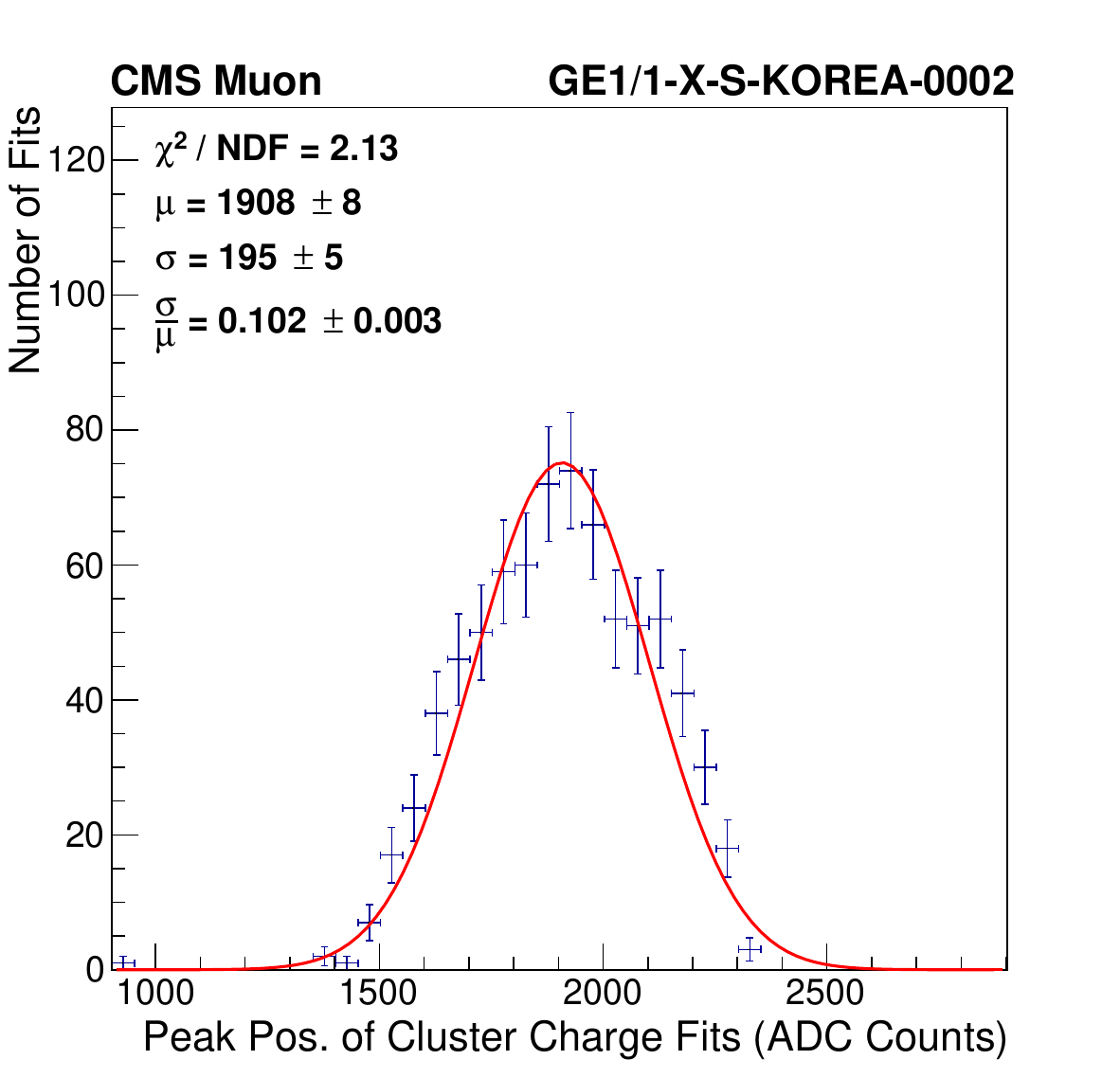}
  }\\
  \subfloat{
    \includegraphics[width=0.45\columnwidth]{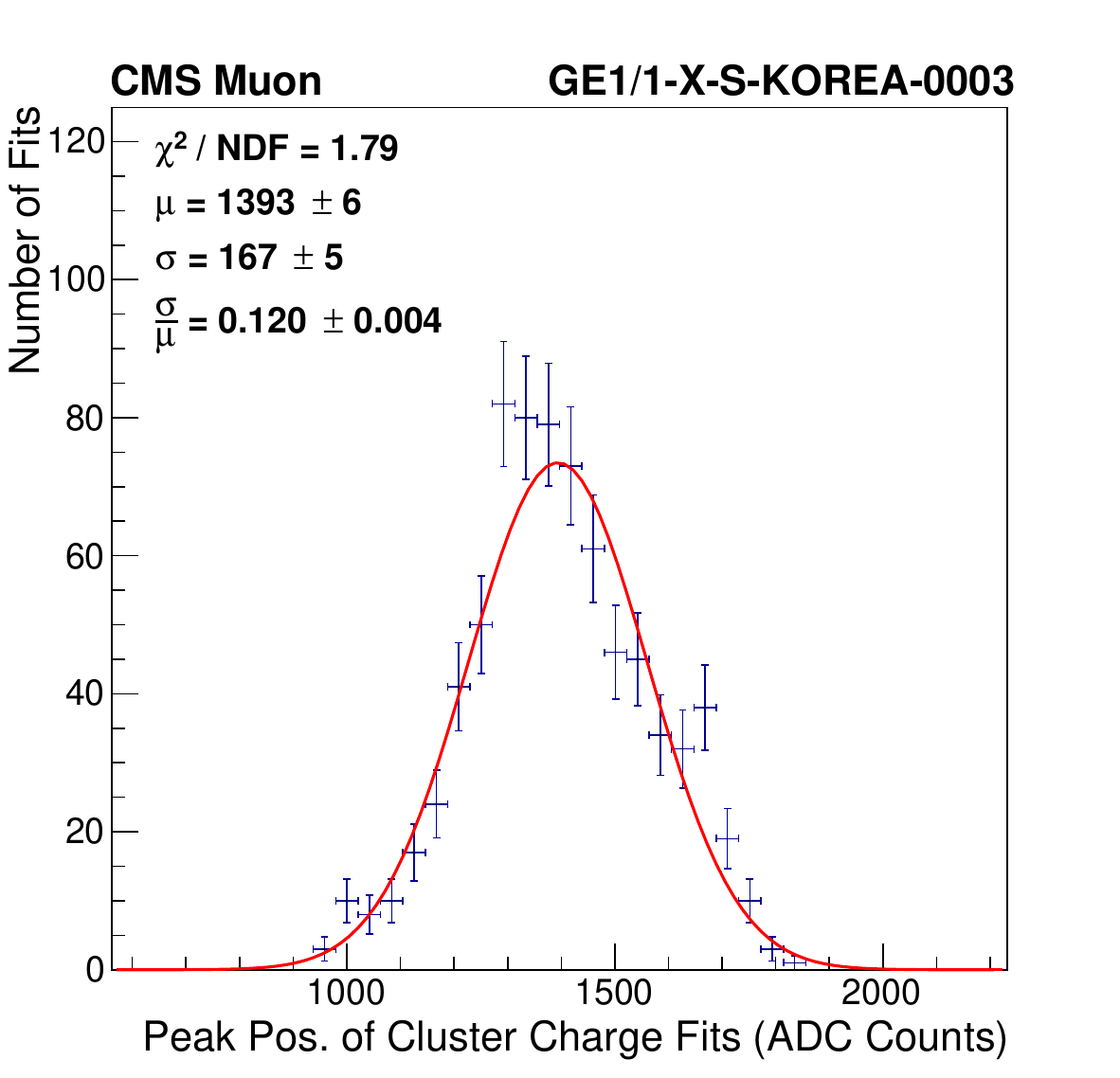}
  }
  \subfloat{
    \includegraphics[width=0.45\columnwidth]{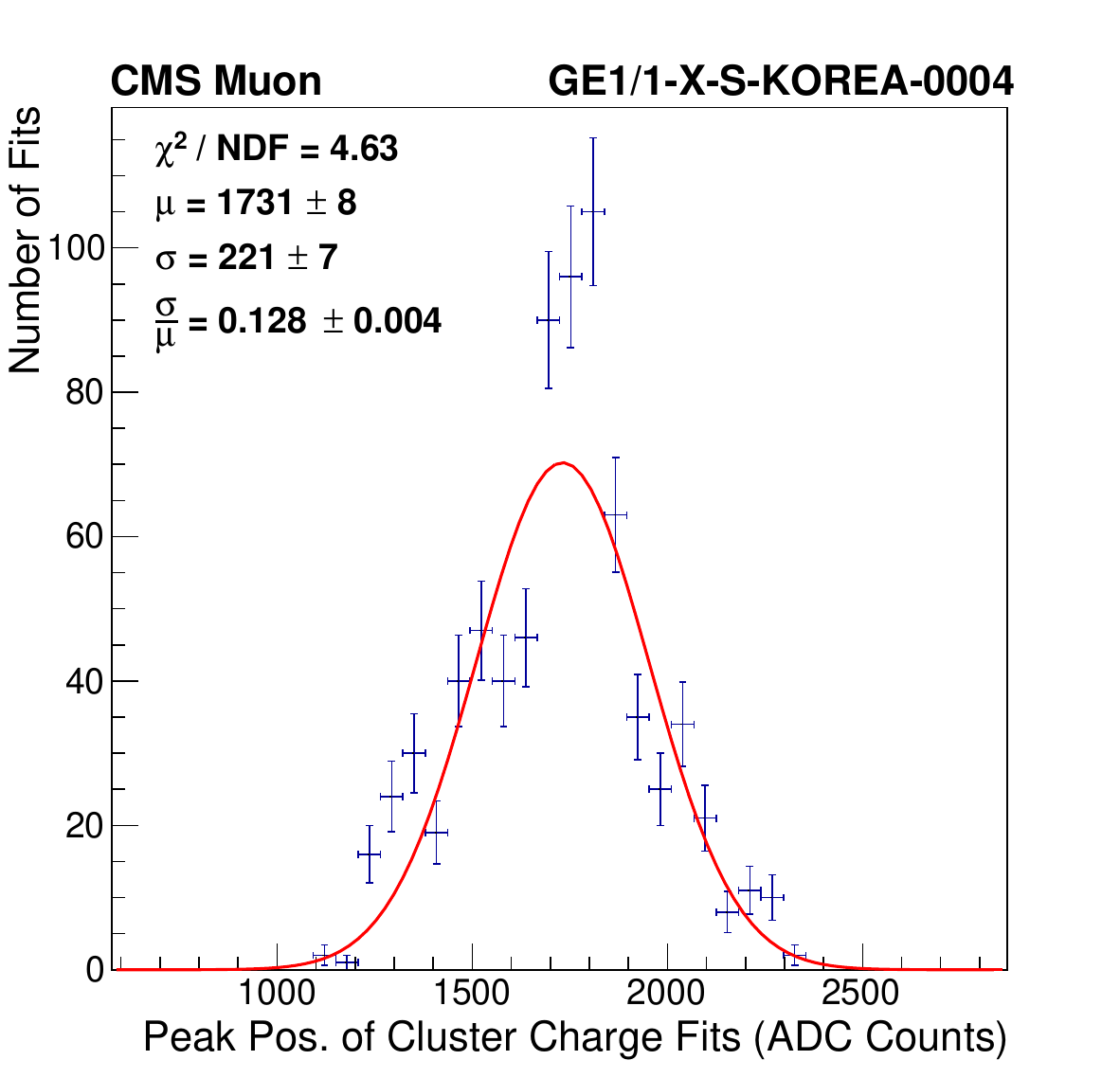}
  }
  \caption[Response uniformity]{
      Peak ADC distributions of CMS GE1/1 short-type detectors with Mecaro foils.
      $\sigma/\mu$ represents the response uniformity of a detector. 
      The response uniformities of the four detectors lie in the range \SIrange{10.2}{16.2}{\percent}, consistent with the uniformities of detectors with CERN foils.
    }\label{fig:gain_uniformity}
\end{figure}

Fig.~\ref{fig:2d_gain_uniformity} illustrates the two-dimensional distributions of this peak ADC value obtained from the four readout strips fit.
Since the GE1/1 detector lacks internal pillars fixing the drift board and readout board, it exhibits a tendency for the center of the detector to warp. 
Consequently, the center of detector shows lower gains, while the edges display higher gains.
This tendency, due to mechanical characteristic of the GE1/1 detector structure, is similarly observed in GE1/1 detectors assembled with CERN foils \cite{ABBAS2022166716}.
Fig.~\ref{fig:gain_uniformity} presents a one-dimensional representation of the peak ADC distribution from Fig.~\ref{fig:2d_gain_uniformity}.
The distributions were fit again with a Gaussian function to obtain mean ($\mu$) and variance ($\sigma$).
Response uniformity, $\sigma/\mu$, of the four detectors are in the range \SIrange{10.2}{16.2}{\percent}, consistent with the uniformities of detectors with CERN foils \cite{VENDITTI2018}.

\textbf{Rate capability}: 
% It is important to measure the rate capability of GEM detectors with Mecaro foils, as high particle rate capability is an advantage of GEM detectors and a reason why GEM technology has been selected for the CMS upgrade.
%
To estimate the rate capability of GEM detectors with Mecaro foils, the effective gas gain of the detector was measured as a function of x-ray flux.
%
%The gain of the detector was measured as discussed in the effective gain measurement.
%
The gain was measured as discussed previously.
X-ray flux was tuned by varying the current of the x-ray tube and the number of Cu attenuators.
To control the area of x-ray exposure, a brass collimator with a \SI{2}{\milli\meter} diameter 
% added word aperture
aperture was used.
Fast preamplifier electronics were utilized to avoid  pileup of the signals.
No pileup was observed up to a signal rate of \SI{5E5}{\hertz} as determined by the linearity of signal rate versus x-ray tube current.
Above that value, the signal rate was estimated via extrapolation based on the linear relationship between the x-ray flux and x-ray tube current, and the effect of the Cu attenuators.

\begin{figure}[hbt]
  \centering
  \includegraphics[width=0.8\columnwidth]{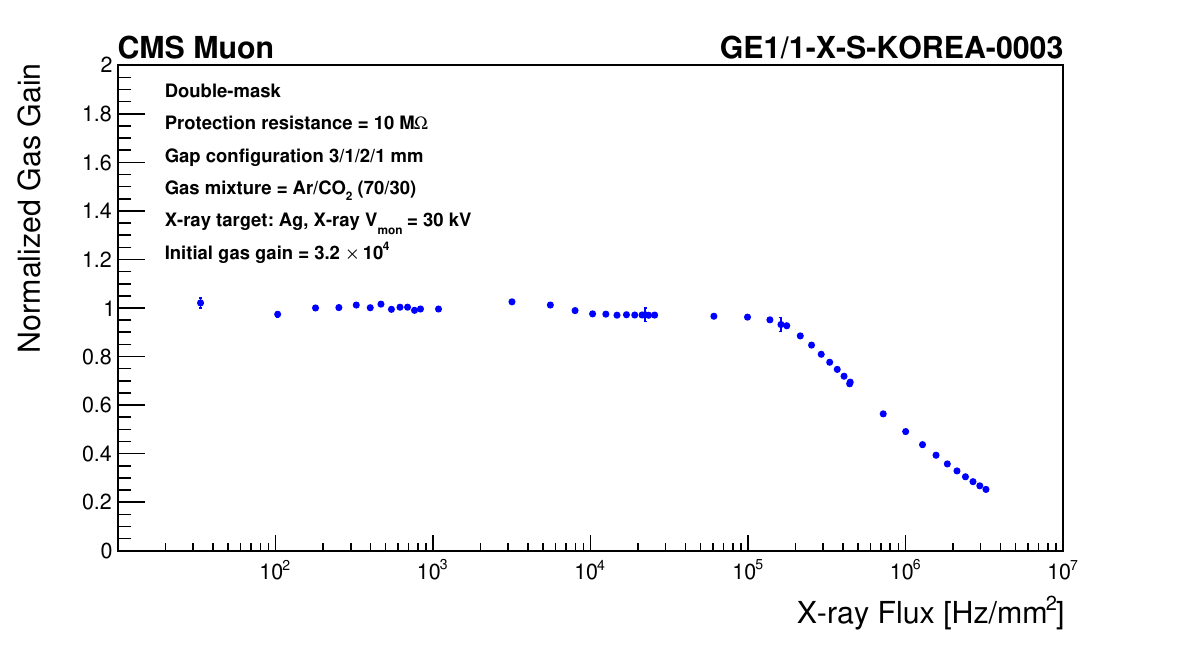}
%  \caption[Rate capability]{Normalized gain curves of CMS GE1/1 short-type detector assembled with Mecaro foils as a function of x-ray flux. The gain remained stable for x-ray fluxes up to \SI{1e5}{\hertz\per\milli\meter\squared}.}
  \caption[Rate capability]{Normalized gas gain versus x-ray flux for a CMS GE1/1 short-type detector assembled with Mecaro foils. The gain remains stable for x-ray fluxes up to \SI{1e5}{\hertz\per\milli\meter\squared}.}
\label{fig:flux_capa}
\end{figure}

% Fig.~\ref{fig:flux_capa} presents the obtained results.
%
The normalized gas gain versus x-ray flux for a CMS GE1/1 short-type detector assembled with Mecaro foils is shown in Fig.~\ref{fig:flux_capa}.
The gain 
% remained 
is
stable for x-ray fluxes up to \SI{1e5}{\hertz\per\milli\meter\squared}.  
%due to referee 2
%The gain drop above that flux is due to the effect of protection resistors in the HV circuit. 
Beyond that flux, no characteristic change is observed in the gain behavior of the GEM detector, which typically follows a pattern of increase and subsequent decrease due to variations in extraction efficiency and gas gain prompted by space charge; instead, the gain decreases \cite{SAULI20162}.
This phenomenon is attributed to a voltage drop originating from the protection resistors positioned as shown in Figure.~\ref{fig:hv_supply}. 
When the detectors are exposed to very high x-ray flux, the current flowing through the GEM foils becomes significant, thereby inducing voltage drops across the resistors and reducing the voltage applied to the GEM foils.
To verify this behavior, detectors with $10 \times 10$ \si{\centi\meter\squared} GEM foils, fabricated in the same way as larger-sized GE1/1 foils, were equipped with small protection resistors.
The flux capability measurement showed the characteristic change in normalized gain at high fluxes.
%
% I surmised that the next sentence is justified. (confirmed by Inseok)
%
The GE1/1 detector with Mecaro foils has comparable flux capability as those fabricated with CERN foils.

%
%Regarding the gain drop owing to the accumulation of space charges, the measured capability is sufficient for the CMS upgrade.
%
Because particle flux varies rapidly with angle from the collision axis, 
the foils of the ME0 detectors will be radially segmented to equalize the gain drops in the high voltage sectors \cite{Bianco_2022}. 
%
% Note that flux is about constant with rapidity, but varies rapidly with angle.
%
%Considering the extreme rapidity dependence of the particle flux, the foil of the ME0 detector will be radially segmented to equalize the gain drop in each high voltage sector in consideration of the voltage drop in protective resistors \cite{Bianco_2022}.
%
% Other conventional applications are also feasible.
% Above statement is vague and not needed for this paper.
% For discussion below include reference to Sauli's seminal paper; 
%Fig. 43 of NIMA 805, 2-24 (https://doi.org/10.1016/j.nima.2015.07.060
%
%Fig.~\ref{fig:flux_capa} does not demonstrate the characteristic increase and decrease in the effective gain of GEM detectors from the accumulation of space charges at high flux, owing to the voltage drop at high protective resistors.
%The CMS GE1/1 type GEM foils are equipped with \SI{10}{\mega\ohm} surface mounted resistors for protection.
%High voltage is applied to active areas of GEM foils via the resistors.
%When the detectors are exposed to very high x-ray flux, the current flowing through GEM foils becomes significant, thereby inducing voltage drops at the resistors.
%Therefore, the voltage applied to GEM foils is reduced in this region. With \TENTEN GEM foil

%\textbf{Discharge probability and robustness to discharge}: 
\textbf{Discharge probability and robustness after discharge}:
%
%Because discharges may damage GEM foil itself or readout electronics, %detectors with low discharge probability are preferred.
%
Because discharges may damage the GEM foil or readout electronics, the discharge probability should be minimized.
%
%Although the triple GEM detector can achieve sufficient gain while maintaining low probability via the step-by-step amplification, the imperfection of the hole development may induce a higher probability of discharges.
%
Since imperfections in the structure of holes in the GEM foil might induce discharges, it is important to verify that a detector with Korean foils exhibits a discharge probability as low as those of detectors with CERN foils.
%
%Therefore, it is important to verify that the GEM detector with Korean foils also exhibit a discharge probability that is as low as that of the detectors with CERN foils.
%
%
%Discharges were induced by \SI{5.5}{\mega\electronvolt} $\alpha$-particles from $^{241}$Am.
%To pass $\alpha$-particles inside the detector, holes with \SI{5}{\milli\meter} diameters were drilled into the drift plate.
%The holes were covered by plastic wraps to prevent gas from leaking.
%Discharges were detected by an electric wire loop around the high voltage line to GEM foil.
%When a discharge occurs, the electromagnetic field induced by the abrupt change in current induces a current in the electric wire.
%The presence of current in the wire loop is used to count the number of discharges. 

Discharges were induced by \SI{5.5}{\mega\electronvolt} $\alpha$-particles from $^{241}$Am that entered the detector
through \SI{5}{\milli\meter} diameter holes drilled into the drift plate.
The holes were covered by plastic to prevent gas from leaking.
Discharges were detected by the current pulse induced in a wire loop surrounding the high voltage line.
%
% How was N_alpha determined? (See below and added statement.)
%
%To induce frequent discharges, the measurement was performed at higher gains than the normal operating gain.
%The discharge probability at the normal operating gain was estimated by extrapolation. For the extrapolation, an exponential function was fitted to the data adopting a maximum likelihood estimation with a Poission likelihood. 
%
To increase the frequency of discharges, the gain was set above the value used in normal operation. 
The discharge probability at normal gain was estimated by extrapolation. 
%For the extrapolation, an exponential function was fitted to the data adopting a maximum likelihood estimation with a Poission likelihood. 

% In the figure, are the vertical error bars on each point statistical? (Yes, I think that is the case.)
%
%Fig.~\ref{fig:discharge_prob} presents the probability and its 68\% confidence interval as a function of the effective gas gain of CMS GE1/1 short-type detectors with Mecaro foils.
%The discharge probabilities are calculated as $\frac{N_{discharge}}{N_{\alpha}}$.  The uncertainties are calculated by the Feldman-Cousins approach \cite{PhysRevD.57.3873} and solely presented for legibility.
%At the gain of \num{1e4}, the probability was estimated as $2.6_{-0.8}^{+1.1}\times10^{-9}$. The probability is consistent with the probability of the detectors with CERN foil \cite{Colaleo:2021453}.

\begin{figure}[hbt]
  \centering
  \includegraphics[width=0.8\columnwidth]{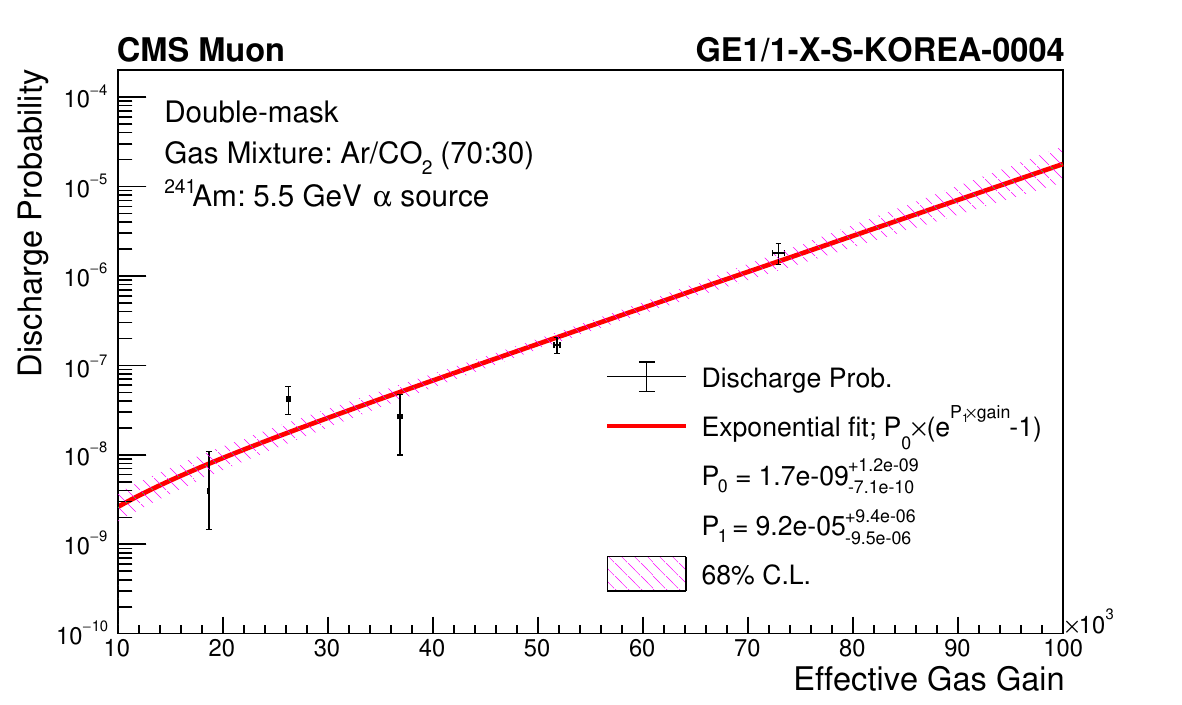}
  %\caption[Discharge probability]{Discharge probability and its 68\% confidence interval as a function of the effective gas gain of the CMS GE1/1 short-type detector assembled with Mecaro foils. The discharges were induced by \SI{5.5}{\mega\electronvolt} $\alpha$-particles from $^{241}$Am.}
  \caption[Discharge probability]{Discharge probability per incident alpha particle and its 68\% confidence interval as a function of effective gas gain for of a GE1/1 short-type detector assembled with Mecaro foils. The discharges were induced by \SI{5.5}{\mega\electronvolt} $\alpha$-particles from $^{241}$Am.}
  \label{fig:discharge_prob}
\end{figure}

Fig.~\ref{fig:discharge_prob} presents the measured discharge probability per incident alpha particle, $\frac{N_{discharge}}{N_{\alpha}}$, as a function of effective gas gain for a GE1/1 short-type detector with Mecaro foils. $N_{\alpha}$ was measured by the number of current pulses from the GEM readout strips.
The uncertainties shown by vertical bars are calculated by the Feldman-Cousins approach \cite{PhysRevD.57.3873}.
% 
% and solely presented for legibility. (This is implicit.)
%
% included statement above about N_alpha being measured by current pulse from the GEM readout strips.
%
The data were fit with an exponential function using Poisson maximum likelihood. 
At the reference gain of \num{1e4}, the discharge probability is $2.6_{-0.8}^{+1.1}\times10^{-9}$, consistent with that of detectors with CERN foils \cite{Colaleo:2021453}.

%The uncertainty band (displayed in the figure) was calculated using the Feldman-Cousins approach \cite{PhysRevD.57.3873}.

%The effective gas gain curves and energy spectra were measured before and after the discharge probability measurement to estimate the robustness of GEM foils to discharges.
%To calculate the energy resolutions, the energy spectra were fitted with two Gaussian functions and a 5th order polynomial function to obtain the full width at half maximum of the main peak.
%As illustrated in Fig.~\ref{fig:discharge_hardness_gain} and Fig.~\ref{fig:discharge_hardness_adc}, the effective gas gain and energy resolution did not degrade even after the detector had suffered at least 229 discharges during the discharge probability measurement.
%Most of the discharges occurred at the extreme high gain.

\begin{figure}[hbt]
  \centering
  \includegraphics[width=0.8\columnwidth]{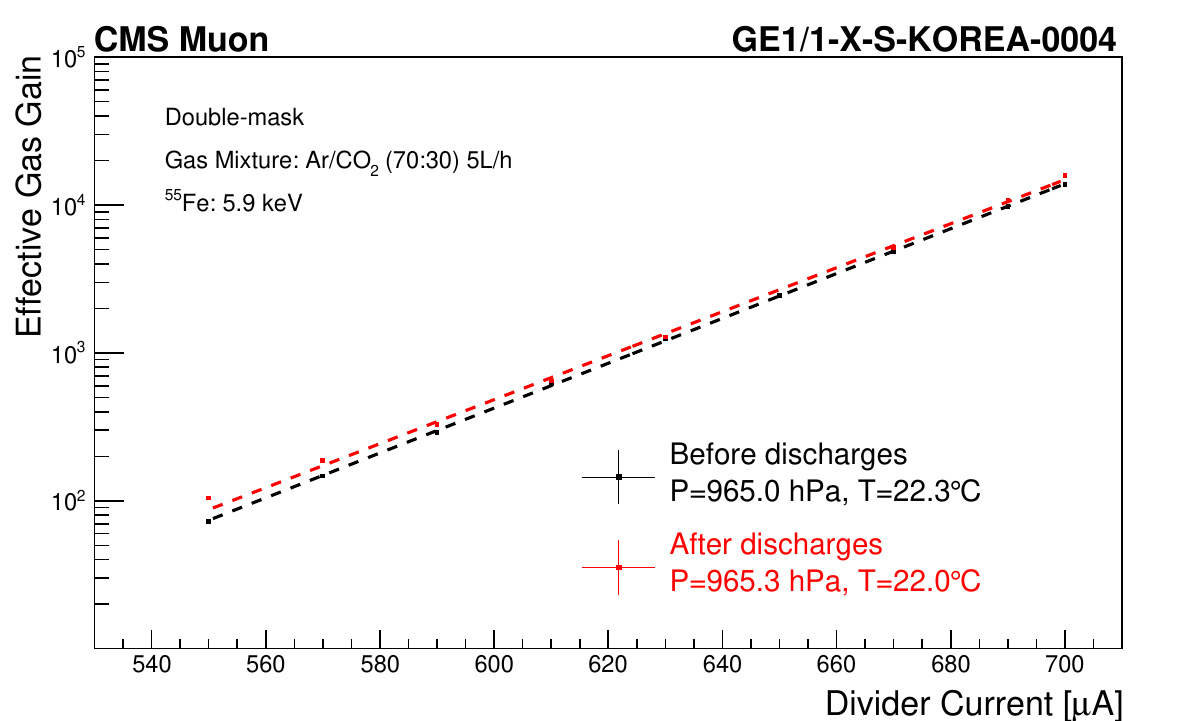}
  \caption[Hardness to discharges]{Gain versus HV divider current for a GE1/1 detector with Mecaro foils before and after operation with discharges.}
  \label{fig:discharge_hardness_gain}
\end{figure}

To assess the robustness of the GEM foils to discharges, effective gas gain and energy resolution were compared before and after operation with discharges. 
There were 229 discharges, most of which occurred at the highest gains.
The measurement method for effective gas gain was discussed previously with examples shown in
Fig.~\ref{fig:gas_gain}. Measurements of gain versus HV divider current for the same detector before and after operation with discharges are shown in Fig.~\ref{fig:discharge_hardness_gain}. There is no degradation in gain.

\begin{figure}[hbt]
  \centering
  \includegraphics[width=0.8\columnwidth]{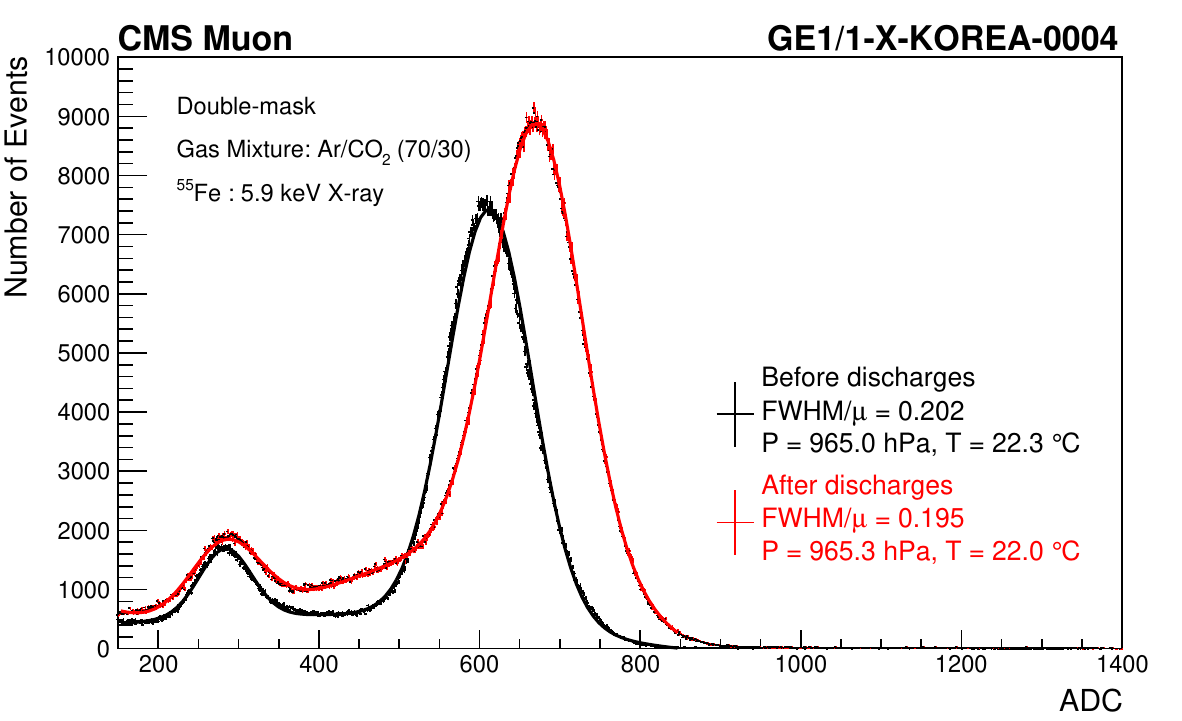}
  \caption[Hardness to discharges]{Energy spectra of x-rays from a $^{55}$Fe source irradiating a GE1/1 detector with Mecaro foils before and after operation with discharges.}
  \label{fig:discharge_hardness_adc}
  \end{figure}

The energy resolution was determined using the 5.9 keV X-ray produced with a $^{55}$Fe source.
The energy spectra, shown in Fig.~\ref{fig:discharge_hardness_adc}, were fit with two Gaussian functions and a 5th order polynomial to obtain the full width at half maximum of the main peak.
The width did not increase after operation with discharges.

Before and after the discharge experiment, \SI{9.5}{\percent} gain fluctuation was observed. GEM detectors typically experience changes of around \SI{10}{\percent} in gain even under normal conditions without specific events such as discharges. 
This could be due to unknown effects of environmental fluctuation or the charging-up phenomenon.
Nevertheless, identifying the precise source of this gain variation poses a formidable challenge.
Notably, the detector's performance did not deteriorate after the discharges. 

\textbf{Aging phenomena}: 
%To ensure the reliable operation of the detectors for at least ten years in the harsh HL-LHC environment, unprecedented durability against classical aging is required.
%Although the GEM detector is known to be robust to classical aging \cite{GUIRL2002263}, background flux at the region where the ME0 station will be installed is extremely high, resulting in accumulated charge of \SI{7.93}{\coulomb\per\centi\meter\squared} for ME0 detectors to collect over ten years of their operation.
%As the detailed design of the detector located in front of the ME0 station progressed, the durability required for ME0 increased significantly \cite{Bianco_2022}.
%
%
The well-established robustness of GEM detectors against classical aging \cite{GUIRL2002263} is important for long-term, reliable operation in the harsh HL-LHC environment. At the time of the studies reported here, the background flux for an ME0 detector was expected to result in an accumulated charge of \SI{283}{\milli\coulomb\per\centi\meter\squared} over ten years of operation.
%
%. Check that statement below is correct.
%
The current estimate of the accumulated charge is about thirty times as large and the ME0 HV sectorization has been modified to increase rate capability \cite{Bianco_2022}. 
%

%To evaluate the aging phenomena of the detector, detector properties, such as the effective gas gain, were monitored while exposed to high flux radiation.
%The first GE1/1 detector assembled with Korean foils was installed in GIF++ on Jan., 2018.
%The initial gain of the detectors was adjusted to be $\num{2e4}$.
%Then the detector was exposed to $\SI{662}{\kilo\electronvolt}$ $\gamma$-ray emitted from $\SI{14.1}{\tera\becquerel}$ (2015).
%While exposed to $\gamma$-ray, $I_{RO}$ was measured to calculate the normalize gas gain and the collected charge.
%Gain fluctuation owing to the change in environmental variables was corrected.
%Further details can be found in \cite{FALLAVOLLITA2019427}. 

To evaluate the effects of aging, properties of detectors with Korean foils were monitored during exposure to high-flux radiation. 
The first GE1/1 detector assembled with Korean foils was installed in GIF++ in January, 2018.
The initial gain was adjusted to be $\num{2e4}$ and
the detector was exposed to $\SI{662}{\kilo\electronvolt}$ $\gamma$-rays emitted from a $\SI{14.1}{\tera\becquerel}$ $^{137}$Cs source whose radioactivity was calibrated in 2015.
%
%. what is the date 2015 for?
% -> the radioactivity of Cs source was calibrated in 2015
%
During exposure, $I_{RO}$ was measured to calculate the normalized gas gain and the collected charge.
Gain fluctuation from environmental variation was 
suppressed 
as described in \cite{FALLAVOLLITA2019427}. 
%
% I think "removed" is too strong because there are still variation from environmental fluctuation even after PT correction. 
% It's hard to remove all the effect of the environmental 
%

\begin{figure}[hbt]
  \centering
  \includegraphics[width=0.6\columnwidth]{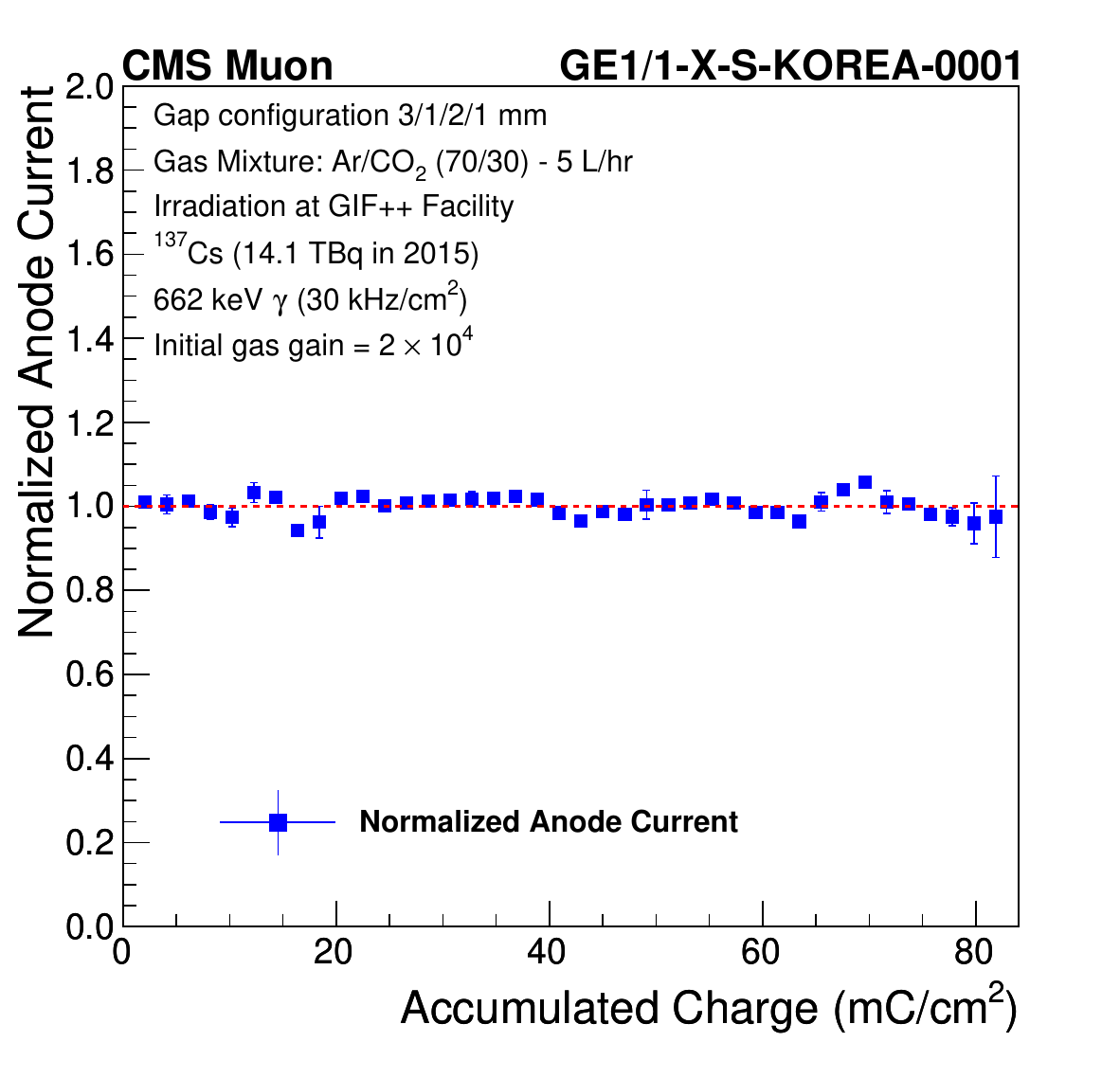}
%  \caption[Aging phenomena]{Normalized gas gain as a function of charge collected by readout to evaluate the aging phenomena.}
  \caption[Aging phenomena]{Normalized gas gain as a function of charge collected by readout to test for evidence of aging.}
  \label{fig:aging}
\end{figure}

%
%Fig. \ref{fig:aging} presents the normalized gas gain as a function of charge collected by readout.
%No degradation of detector performance was observed up to a collected charge of \SI{82}{\milli\coulomb\per\centi\meter\squared}.
%The result safely satisfies the requirement of GE2/1 detectors, as \SI{82}{\milli\coulomb\per\centi\meter\squared} corresponds to the collected charge expected to be obtained from the operation of GE2/1 detectors for 273 years under HL-LHC condition.
%
Fig. \ref{fig:aging} presents the normalized gas gain as a function of charge collected by readout.
No degradation of detector performance was observed up to a collected charge of \SI{82}{\milli\coulomb\per\centi\meter\squared} that corresponds to the value expected for GE2/1 detectors after operation for 273 years under HL-LHC conditions.

%However another aging study is being conducted because ME0 detectors require significantly robust durability.
%This study adopts an x-ray generator that utilizes an Ag transmission target as the radiation source instead of the $^{137}$Cs to increase charge collection rate.
%Because the interaction cross section of the lower energy x-ray is significantly higher than that of the $\gamma$-ray from $^{137}$Cs, a faster charge collection rate is achievable, although the intensity of the x-ray generator is lower than the $^{137}$Cs \cite{FALLAVOLLITA2019427}.
%This aging study is being performed with the CMS GE2/1 M2 type detector assembled with double-segmented GEM foils, which are novel versions of foils used in controlling the discharge propagation to adjacent foils \cite{Jeremie_MPGD19}, instead of the CMS GE1/1 short detector assembled with single-segmented foils. 

Because of the much larger accumulated charge expected for ME0 compared to GE1/1 and GE2/1, new measurements are in progress to study aging behavior with much larger fluxes than obtained in 2018. The current measurements utilize an x-ray generator with a Ag transmission target as the radiation source instead of the $^{137}$Cs source.
Because the interaction cross section of the lower energy x-ray is significantly higher than that of the $\gamma$-ray from $^{137}$Cs, a faster charge collection rate is achievable even though the intensity of the x-ray generator is lower than that of the $^{137}$Cs source \cite{FALLAVOLLITA2019427}.
 The detector under test is a GE2/1 M2 detector, which is equipped with double-segmented foils. 
 The foil is divided into several sectors, encompassing an area of approximately \SI{10}{\centi\meter\squared}, and each sector is electrically 
 %insulated 
 isolated
 through resistive coupling on both the top and bottom sides of the GEM foil.
 This configuration effectively inhibits the propagation of discharges from one foil to the adjacent foil \cite{Jeremie_MPGD19}.
 In contrast, the GE1/1 detectors use single-segmented foils as described at the beginning of Section \ref{sec::validation}. 
\section{Mass production results and plan}
%Owing to the successful completion of the QC and QA  of the GEM foils provided by the new vendor the KCMS and Mecaro consortium are currently in mass production for the CMS GE2/1 upgrade.
%The consortium is expected to produce foils for four modules of GE2/1 and ME0.
%The mass production of GE2/1 foils has begun.
%The produced foils are being sent to CERN for detector assembly after rigorous optical and electrical quality inspection.
%At the end of the mass production the collaboration will provide a reliable vendor for large are GEM foils. 
The successful completion of QC and QA of GEM foils provided by the new vendor enabled the KCMS collaboration to begin mass production of these foils. 
From May 2021 to September 2022, KCMS produced GE2/1 foils, of which 292 foils passed the QC criteria and are being used in the assembly of GE2/1 detectors. 
The 292 foils correspond to \SI{64}{\percent} of the number requested to be built by KCMS. Unfortunately, mass production was suspended because the KCMS-Mecaro consortium ended. The production facility is being relocated with completion expected by mid-2023, after which KCMS will start mass production of ME0 foils.

\section{Summary}
%We have validated the performances of the GEM foils with an area of approximately  \GEoneoneArea produced by a new vendor, the KCMS and Mecaro consortium, which adopts the double-mask technique for the GE2/1 and ME0 phase-\Romannum{2} of CMS.
We have validated the performance of GEM foils with an area of approximately  \SI{0.6}{\meter\squared} produced by a new vendor, the KCMS and Mecaro consortium, employing the double-mask technique for the GE2/1 and ME0 detectors for the Phase-2 upgrade of CMS.
After verifying that the dimension and structure of the produced foils were within the designed range, electrical evaluation of the foil's cleanliness was performed based on the QC methodology developed for the CMS upgrade.
Subsequently, CMS GE1/1 short-type detectors were assembled with the foils.
After the integrity of the assembly was validated, several properties of the detectors were measured.
The detectors assembled with Korean foils were 
%
% as good as 
%
comparable in quality to
the CERN detectors, and they satisfied the requirements of the CMS upgrade in terms of the effective gain, response uniformity, rate capability, discharge probability, and robustness to discharges.
Evaluation of hardness to classical aging was studied and
no degradation of the detector performance was observed up to a charge collection of \SI{82}{\milli\coulomb\per\centi\meter\squared}.
A new aging study for the ME0 requirement is in progress.
The successful QC and QA results enabled the KCMS collaboration to begin mass production of GE2/1 foils.
Of these foils, 292 passed the QC criteria and are being used in the assembly of GE2/1 detectors.
%The equipment transfer is underway as the consortium between the KCMS and the Mecaro has ended.
Since the consortium between the KCMS collaboration and Mecaro has ended, 
equipment is being transferred to KCMS for 
mass production of ME0 foils.

\section*{Acknowledgments}
%We would like to thank Dr. Rui De Oliveira (CERN) for providing invaluable assistance during foil development.
%We are grateful to Taeseong Jeong (Mecaro) and Inseung Jeong (Mecaro) for their contribution to the production of foils.
We thank Dr.~Alexander J. G. Lunt (CERN) for taking SEM images of the foils and Dr.~Matthew Posik (Temple University) for measuring the uniformity of hole diameters.
We gratefully acknowledge financial support from FRS-FNRS (Belgium),
FWO-Flanders (Belgium), BSF-MES (Bulgaria), MOST and NSFC (China),
BMBF (Germany), CSIR (India), DAE (India), DST (India), UGC (India),
INFN (Italy), NRF (Korea), MoSTR (Sri Lanka), DOE (USA), and NSF
(USA).

%\section*{Reference}
%\bibliographystyle{elsarticle-num}
%\bibliography{CERN,NIM,PRD,Misc}

\begin{thebibliography}{99}
\bibitem{SAULI1997531} F. Sauli, GEM: a new concept for electron amplification in gas detectors, Nucl. Instrum. Methods Phys. Res. A, 386 (1997) 531 - 534, \url{http://www.sciencedirect.com/science/article/pii/S0168900296011722}
\bibitem{Apollinari:2284929} G. Apollinari, et al., High-luminosity large hadron collider (HL-LHC): Technical Design Report V. 0.1, CERN, Geneva, 2017, \url{http://cds.cern.ch/record/2284929}
\bibitem{Colaleo:2021453} CMS Collaboration, CMS TDR, CERN-LHCC-2015-0122015, 2015, \url{https://cds.cern.ch/record/2021453}
\bibitem{Collaboration:2283189} CMS Collaboration, CMS TDR, CERN-LHCC-2017-012, 2017, \url{https://cds.cern.ch/record/2283189}
\bibitem{LHC_Operation_Schedule} LHC longer term schedule is available at \url{https://lhc-commissioning.web.cern.ch/schedule/LHC-long-term.htm} 
%\bibitem{Schmidt_2016} B. Schmidt, The high-luminosity upgrade of the {LHC}: physics and technology challenges for the accelerator and the experiments, J. Phys.: Conf. Ser., 706 022002, \url{https://doi.org/10.1088%2F1742-6596%2F706%2F2%2F022002} this should be updated. There should be better ref
\bibitem{Lamont_2022} M. Lamont, LHC accelerator: status and perspectives, plenary talk at ICHEP 2022, Bologna, \url{https://agenda.infn.it/event/28874/contributions/171905/} 
\bibitem{Mecaro} Mecaro Co. Ltd., Wonnamsandan-ro, Wonnam-myeon, Eumseong-gun, Chungcheongbuk-do, Republic of Korea, 27721, \url{http://www.mecaro.com} 
\bibitem{Pinto_2009} S.D. Pinto, et al., Progress on large area GEMs, JINST 4, (2009) 12009, \url{https://doi.org/10.1088%252F1748-0221%252F4%252F12%252Fp12009}
%\bibitem{Pubchem_MEA} National Center for Biotechnology Information, PubChem compound summary for CID 700, ethanolamine, available at \url{https://pubchem.ncbi.nlm.nih.gov/compound/Ethanolamine} (accessed on 8th September, 2022)
%\bibitem{Pubchem_EDA} National Center for Biotechnology Information, PubChem compound summary for CID 3301, ethylenediamine, available at \url{https://pubchem.ncbi.nlm.nih.gov/compound/Ethylenediamine} (accessed on 8th September, 2022)
\bibitem{Yoon:2023gyf} I. Yoon, Techniques for mass production of large-sized GEM foil by the Korean CMS group for CMS phase-2 upgrade, JINST 18 (2023) C06010, \url{https://doi.org/10.1088/1748-0221/18/06/C06010}
\bibitem{PFEIFFER201791} D. Pfeiffer, et al., The radiation field in the gamma irradiation facility GIF++ at CERN, Nucl. Instrum. Methods Phys. Res. A, 866 (2017) 91 - 103, \url{http://www.sciencedirect.com/science/article/pii/S0168900217306113}
\bibitem{POSIK201510} M. Posik, B. Surrow, Optical and electrical performance of commercially manufactured large GEM foils, Nucl. Instrum. Methods Phys. Res. A, 802 (2015) 10 - 15, \url{http://www.sciencedirect.com/science/article/pii/S0168900215010001}
\bibitem{Posik_private} M. Posik, Personal communication, (2017)
\bibitem{Abbaneo_2015} D. Abbaneo, et al., Quality control and beam test of {GEM} detectors for future upgrades of the {CMS} muon high rate region at the {LHC}, JINST 10 (2015) C03039, \url{https://doi.org/10.1088%2F1748-0221%2F10%2F03%2Fc03039}
\bibitem{VENDITTI2018} R. Venditti, Production and quality control of the new chambers with GEM technology in the CMS muon system, Nucl. Instrum. Methods Phys. Res. A, 936 (2018) 476 - 478, \url{http://www.sciencedirect.com/science/article/pii/S016890021831581X}
\bibitem{FRENCH2001359} M.J. French et al., Design and results from the APV25, a deep sub-micron CMOS front-end chip for the CMS tracker, Nucl. Instrum. Methods Phys. Res. A, 466 (2001) 359 - 365, \url{http://www.sciencedirect.com/science/article/pii/S0168900201005897}
\bibitem{ABBAS2022166716} M. Abbas et al., Quality control of mass-produced GEM detectors for the CMS GE1/1 muon upgrade, Nucl. Instrum. Methods Phys. Res. A, 1034, (2022) 166716, \url{https://www.sciencedirect.com/science/article/pii/S0168900222002480}
\bibitem{SAULI20162} F. Sauli, The gas electron multiplier (GEM): Operating principles and applications, Nucl. Instrum. Methods Phys. Res. A, 805 (2016) 2 - 24, \url{https://www.sciencedirect.com/science/article/pii/S0168900215008980}
\bibitem{Bianco_2022} M. Bianco et al., High rate capability studies of triple-{GEM} detectors for the {ME}0 upgrade of the {CMS} muon spectrometer, JINST 17 (2022) C02009, \url{https://doi.org/10.1088/1748-0221/17/02/c02009}
\bibitem{PhysRevD.57.3873} G.J. Feldman, R.D. Cousins, Unified approach to the classical statistical analysis of small signals, Phys. Rev. D, 57 (1998) 3873 - 3889, \url{https://link.aps.org/doi/10.1103/PhysRevD.57.3873}
\bibitem{GUIRL2002263} L. Guirl, S. Kane, J. May, J. Miyamoto, I. Shipsey, Nucl. Instrum. Methods Phys. Res. A, 478 (2002) 263-266, \url{https://www.sciencedirect.com/science/article/pii/S0168900201017685}
\bibitem{FALLAVOLLITA2019427} F. Fallavollita, Aging phenomena and discharge probability studies of the triple-GEM detectors for future upgrades of the CMS muon high rate region at the HL-LHC, Nucl. Instrum. Methods Phys. Res. A, 936 (2019) 427 - 429, \url{http://www.sciencedirect.com/science/article/pii/S0168900218314943}
%\bibitem{Jeremie_RD51} J.A. Merlin, Discharge and crosstalk mitigation for the GE21 system in CMS, RD51 collaboration meeting and topical workshop on new horizons in time projection chambers, (2020), \url{https://indico.cern.ch/event/889369/contributions/4040763/} there should be published result.
\bibitem{Jeremie_MPGD19} J.A. Merlin, Study of discharges and their effects in GEM detectors, talk at MicroPattern Gaseous Detectors Conference, 5-10 May, 2019, La Rochelle, France, \url{https://indico.cern.ch/event/757322/contributions/3396501/}
\end{thebibliography}

\end{document}